\documentclass[10pt,aps,preprint,prd,showpacs,nofootinbib,preprintnumbers,a4paper,superscriptaddress]{revtex4-1}
\pdfoutput=1
\usepackage{amsmath}
\usepackage{amssymb}
\usepackage[pdftex]{graphicx}
\usepackage{epstopdf}
\usepackage[caption=false]{subfig}
\usepackage[english]{babel}
\usepackage{amsfonts}
\usepackage{bbold}
\usepackage{bm}
\usepackage{color}
\usepackage{xcolor}
\usepackage{dcolumn}
\usepackage[colorlinks]{hyperref}
\usepackage{cancel}
\usepackage{units}
\usepackage[normalem]{ulem}
\usepackage{slashed}
\usepackage{feynmp}
\usepackage{ulem}
\hypersetup{
pdfauthor={},
pdftitle={strongly coupled hidden sector}
}

\newcommand{\diag}{\ensuremath{\mathrm{diag}}}

\newcommand{\nn}{\nonumber}

\newcommand{\be}{\begin{eqnarray}}
\newcommand{\ee}{\end{eqnarray}}
\catcode`\@=11
\def\lsim{\mathrel{\mathpalette\@versim<}}
\def\gsim{\mathrel{\mathpalette\@versim>}}
\def\@versim#1#2{\vcenter{\offinterlineskip
\ialign{$\m@th#1\hfil##\hfil$\crcr#2\crcr\sim\crcr } }}
\catcode`\@=12

\begin{document}

\title{Gamma-ray Line  from  Nambu-Goldstone Dark Matter \protect\\
in a Scale Invariant Extension of the Standard Model \vspace{4mm}}
\vspace*{20mm}

\author{Jisuke Kubo}
\email{jik@hep.s.kanazawa-u.ac.jp}
\affiliation{Institute for Theoretical Physics, Kanazawa University, Kanazawa 920-1192, Japan}

\author{Kher Sham Lim}
\email{khersham.lim@mpi-hd.mpg.de}

\author{Manfred Lindner \vspace{2mm}}
\email{lindner@mpi-hd.mpg.de}
\affiliation{Max-Planck-Institut f\"{u}r Kernphysik, 69117 Heidelberg, Germany \vspace{5mm}}

\preprint{KANAZAWA-14-05}

\begin{abstract}

A recently proposed  scale invariant extension of the standard model
is modified such that it includes a Dark Matter candidate which can annihilate
into gamma-rays. For that
a non-zero $U(1)_Y$ hypercharge $Q$  
is assigned 
to the fermions
in  a  QCD-like hidden sector.
The Nambu-Goldstone bosons, that arise due to 
dynamical chiral symmetry breaking in the hidden sector,
are cold Dark Matter candidates,
and the extension allows them to annihilate
into two photons, producing a $\gamma$-ray line spectrum.
We find that the $\gamma$-ray line energy
must be  between $0.7$ TeV  and $0.9$ TeV 
with  the velocity-averaged   annihilation
cross section
$10^{-30} \sim \unit[10^{-26}]{cm^3/s}$ 
for $Q=1/3$.
With a non-zero hypercharge $Q$, the hidden sector is no longer
completely dark and can be directly probed 
by collider experiments.

\end{abstract}

\maketitle

\section{Introduction}
With the discovery of the Higgs particle \cite{Aad:2012tfa,Chatrchyan:2012ufa}
the standard model (SM) is now complete.
However, the SM must be extended since it does not contain a Dark Matter (DM) 
candidate and since finite neutrino masses must also be included. 
From a pure theoretical point of view there exist also severe conceptual problems 
and one of them is that the SM cannot explain the origin of its energy scale.
Theoretically, we can imagine a world without any energy scale, but in the real 
world of elementary particles scale invariance is  broken. In the SM the mass term
of the Higgs field is the only term in the Lagrangian that violates (at tree level) 
scale invariance.
Although the SM does not explain the origin of its energy scale,
the measured mass $m_h$  of the Higgs particle seems to suggest
how to go beyond the SM,
because this mass value together with the top quark mass
implies that the SM remains perturbative at least up to the Planck scale
\cite{Holthausen:2011aa,Degrassi:2012ry,Bezrukov:2012sa};
an ultraviolet (UV) completion of the SM is not needed.
Any extension which modifies the high energy behavior of the SM 
should therefore be well motivated (see e.g. \cite{Kubo:2014ova}), 
since it may require a UV completion at lower scales.

Introducing an explicit Higgs mass term in the SM does not only break 
classical scale invariance, but it also leads to another severe issue 
known as the gauge hierarchy problem, namely the quadratic sensitivity 
of quantum corrections to high scales. 
It is therefore tempting to start from classically scale invariant 
theories where the SM scale emerges from dimensional transmutation. 
Various attempts to introduce an energy  scale in this way exist 
in the literature \cite{Coleman:1973jx}-\cite{Chankowski:2014fva}.
Scale invariance is broken at the quantum level even in perturbation 
theory \cite{Callan:1970yg}, but it has been argued that the 
protective features of conformal symmetry are not completely 
destroyed \cite{Bardeen:1995kv}. Specifically logarithmic sensitivities would exist, while 
quadratic divergencies would be absent. 

We follow the idea that the energy scale in a classically
scale invariant theory is generated by D$\chi$SB in a QCD-like hidden 
sector, which is  transmitted via a SM singlet messenger field to the SM sector 
\cite{Hur:2007uz,Hur:2011sv,Heikinheimo:2013fta,Holthausen:2013ota} \footnote{See 
also \cite{Strassler:2006im}.}. So we assume that the fermions in the hidden 
sector are SM singlet and allow the presence of fundamental scalar fields.
In fact, the messenger is assumed to be the simplest possibility, a real SM 
singlet scalar $S$. 
Note that this avoids the well known phenomenological problems of 
technicolor models and the model looks very much like the SM, since 
the hidden sector couples only via the Higgs portal.

The possibility that DM annihilates into $\gamma$-ray lines has recently 
received much attention and we want to discuss this possibility therefore in this paper.
Specifically we consider a simple extension of the above mentioned model, where we 
assign a $U(1)_Y$ hypercharge $Q$ to the hidden sector fermions such that  they are
electrically  charged with a charge $Q$.
Since the coupling is vector-like,  no breaking of 
$U(1)_Y$ is caused by D$\chi$SB in the hidden sector.
As we will see, this non-zero charge makes it possible that
DM particles, which are 
the pseudo Nambu-Goldstone particles
in this model, can be annihilated into two photons, producing a $\gamma$-ray line spectrum. 
Monochromatic $\gamma$-ray lines from DM annihilation  exist 
in other DM models, too
\cite{Bringmann:2007nk,Bertone:2009cb},
and in fact experimental searches for
$\gamma$-ray lines have been 
undertaken with Fermi LAT 
\cite{Ackermann:2012qk,Gustafsson:2013fca} and HESS
\cite{Abramowski:2013ax}
for a wide range of high energies.
We find that the energy
of the $\gamma$-ray line in our model
 lies between $0.7$ TeV and $0.9$ TeV.
 (We are not aiming to explain 
the recent observations of the galactic keV X-ray 
\cite{Bulbul:2014sua,Boyarsky:2014jta} here.)
 The upper limits on the velocity-averaged  annihilation cross section $\langle v\sigma \rangle$ given by
Fermi LAT and HESS constrain the electric charge
$Q$ of the hidden fermions.
We find that 
the $\langle v\sigma \rangle$ is
$10^{-30} \sim \unit[10^{-26}]{cm^3/s}$ 
for $Q=1/3$,
which can well satisfy the experimental constraints
of Fermi LAT and HESS.
Since $\langle v\sigma \rangle$ is proportional to
$Q^4$,  our calculations can be simply 
extended to the case of an arbitrary $Q$.

In this model not only DM  particles but also hidden baryons are stable.
For an arbitrary $Q$ the electric charge of the hidden baryons
are fractionally charged.  Note that the consistent 
range of the DM mass between $0.7$ TeV and $0.9$ TeV
is  independent of $Q$ and hence the scale $\Lambda$ of the hidden sector
is roughly fixed (regardless  of $Q$), which means that the mass of the stable hidden baryons is $\sim 3$ TeV.
Using the fact that  the hidden sector is basically described 
by a scaled-up QCD, we have found that the relic abundance of  
the hidden baryons $\Omega_{hB} h^2$ in the Universe is at most $10^{-4}$,
which is independent of $Q$.
This is sufficiently below the upper bound given in \cite{Dolgov:2013una} and the constraint in the $Q$-DM mass plane given in \cite{Langacker:2011db} is also satisfied. Consequently 
there is  practically  no constraint 
(except for those  from FermiLat and HESS) 
on the fractionally  charged hidden baryons.

Since the hidden sector (strictly speaking it is no longer a hidden sector,
because the fermions are electrically charged)
can now communicate through gauge boson exchange
(photon and $Z$ boson) with the SM sector,
the hidden sector could be produced at the ILC.
We postpone these interesting processes for future studies,
as our main priority in this paper is to find a prescription  to obtain gauge invariant
amplitudes. This is because we approximate the strongly
coupled QCD-like  sector by 
the Nambu-Jona-Lasinio model (NJL) \cite{Nambu:1960xd,Nambu:1961tp}
(see \cite{Klevansky:1992qe,Hatsuda:1994pi} for reviews),
which is defined with a finite cutoff $\Lambda$ that violates gauge invariance.
To overcome this problem, we propose least subtraction procedure.
In the NJL model the cutoff $\Lambda$ is a physical parameter 
and a finite  $\Lambda$ is  essential
to describe effectively D$\chi$SB.  We therefore stress that
we keep the subtraction terms to the minimum necessary.

\section{The model}
We consider an extension of  the model studied in 
\cite{Hur:2007uz,Hur:2011sv,Heikinheimo:2013fta,Holthausen:2013ota} 
which consists of a hidden QCD-like sector coupled via a real singlet scalar $S$ to  the SM.
The fermion $\psi$ in the hidden sector belongs to the fundamental representation
of the hidden gauge group $SU(3)_H$.
With this setting 
D$\chi$SB in the hidden sector
does not break  the SM gauge symmetries,
thereby  avoiding the FCNC problem.
This is one of the
main differences to technicolor model.
If we further assume that the Yukawa coupling $\bar{\psi}\psi S$ 
respects $SU(N_f)_V$ flavor symmetry, there is only one coupling
constant $y$ for the Yukawa coupling, so that in the hidden sector
there are only two independent parameters;
the gauge coupling constant $g_H$ and the Yukawa coupling $y$.

In extending the model
we impose that   neither the SM gauge symmetry nor
the $SU(N_f)_V$ flavor symmetry is broken 
in the hidden sector.
If we further impose that the matter content remains unchanged,
then there is a unique possibility for the extension that 
the hidden (Dirac) fermion carries a common $U(1)_Y$ charge $Q$
\footnote{
The new gauge coupling contributes only to
$\Pi_{YY}$ of the gauge boson self-energy diagrams so that
 the $S, T, U$ parameters remain unchanged.}.
This implies that the hidden sector Lagrangian of the extended model 
is written as 
\begin{align}
{\cal L}_{\rm H}
&=-\frac{1}{2}\mbox{Tr}~F^2+
\mbox{Tr}~\bar{\psi}(i\gamma^\mu \partial_\mu +
g \gamma^\mu G_\mu +g'  Q \gamma^\mu B_\mu-
y S)\psi~,
\label{eq:LH}
\end{align}
where $G_\mu$ is the gauge field for the hidden QCD,
and $B$ is the $U(1)_Y $ gauge field.
The trace is taken over the flavor as well as the color indices.
The ${\cal L}_{\mathrm{SM}+S}$ part of the total Lagrangian ${\cal L}_T ={\cal L}_{\rm H}+{\cal L}_{\mathrm{SM}+S} $, which  contains the 
SM gauge and Yukawa  interactions
along with the scalar potential
\be
V_{\mathrm{SM}+S}
&=&
\lambda_H ( H^\dag H)^2+
\frac{1}{4}\lambda_S S^4
-\frac{1}{2}\lambda_{HS}S^2(H^\dag H)~,
\label{eq:VSM}
\ee
is unchanged \footnote{This classically scale invariant 
 model is perturbatively renormalizable, and 
 the Green's functions are infrared finite
 \cite{Lowenstein:1975rf,Poggio:1976qr}.}. $H^T=( H^+ ~,~(h+iG)\sqrt{2}  )$ is the SM Higgs doublet field,
with $H^+$ and $G$ as the would-be Nambu-Goldstone fields.

Here we follow \cite{Holthausen:2013ota} 
in which the NJL model
is used to describe D$\chi$SB in the hidden sector,
restricting ourselves to $N_c=N_f=3$, because
in this case the NJL model parameters, up-to an overall scale,
can be fixed from  hadron physics 
\cite{Kunihiro:1983ej,Hatsuda:1994pi}.
So at low energy we replace the Lagrangian ${\cal L}_H$ by
\be
{\cal L}_{\rm NJL}&=&\mbox{Tr}~\bar{\psi}(i\gamma^\mu\partial_\mu 
+g' Q \gamma^\mu B_\mu-y S)\psi+2G~\mbox{Tr} ~\Phi^\dag \Phi
+G_D~(\det \Phi+h.c.)~,
\label{eq:NJL10}
\ee
where
\be
B_\mu &=&\cos \theta_W A_\mu-\sin\theta_W Z_\mu~,~
g'=e/\cos \theta_W ~,\\
\Phi_{ij}&=& \bar{\psi}_i(1-\gamma_5)\psi_j=
\frac{1}{2}
\lambda_{ji}^a \mbox{Tr}~\bar{\psi}\lambda^a(1-\gamma_5)\psi~,
\ee
and $\lambda^a$ are the Gell-Mann matrices with
$\lambda^0=\sqrt{2/3}~{\bf 1}$.
The last term in (\ref{eq:NJL10}), which exhibits
a six fermi interaction, is present due to chiral anomaly
of the axial $U(1)_A$.
The  chiral symmetry $U(3)_L\times U(3)_R$
is explicitly broken down to its diagonal subgroup $U(3)_V
=SU(3)_F\times U(1)_V$
by the Yukawa coupling with the singlet $S$.
To deal with the non-renormalizable Lagrangian (\ref{eq:NJL10}) 
 we have used in \cite{Holthausen:2013ota}
 a  self-consistent mean-field approximation which has been intensely studied by Hatsuda and Kunihiro
\cite{Kunihiro:1983ej,Hatsuda:1994pi} for hadron physics. 
The effective Lagrangian ${\cal L}_{\rm NJL}$ has
three dimensional parameters
$G,G_D$ and the cutoff $\Lambda$, which have canonical dimensions of 
$-2$, $-5$ and $1$,  respectively.
Since the original Lagrangian ${\cal L}_{H}$ has only one independent scale,
the parameters $G,G_D$ and  $\Lambda$
are not independent.
 We obtain the NJL parameters for the hidden QCD from the upscaling of actual values of $G,G_D$ and the cutoff $\Lambda$ from QCD hadron physics.
 That is, we assume that the dimensionless combinations
 \be
G^{1/2} \Lambda &=&2.0~,~(-G_D)^{1/5} \Lambda =2.1~,
\label{G-GD}
\ee
which are  satisfied for hadrons, remain unchanged for
a higher scale of $\Lambda$ \cite{Holthausen:2013ota}.

In what follows we briefly outline the approximation method
\cite{Kunihiro:1983ej,Hatsuda:1994pi}.
One assumes that the dynamics of the theory creates a chiral symmetry breaking condensate 
\be
\langle0 \vert \bar{\psi}_i \psi_j \vert 0 \rangle  &=&
-\frac{1}{4G}  \diag(\sigma,\sigma,\sigma)~,
\label{eq:bbb}
\ee
which is treated as a classical field $\sigma$. 
The vacuum $| 0 \rangle $ is defined by the annihilation operator
of the constituent fermion $\psi$ in the background of the mean fields.
We  restrict our discussion (in a more complete treatment, one may add terms involving $\eta$ or $\rho$ mesons) to the mean fields collected in 
 \be
 \varphi &  \equiv & \langle0 \vert\bar{\psi}(1-\gamma_5)\lambda^a \psi
 \vert 0 \rangle=
-\frac{1}{4G}  \left(\diag(\sigma,\sigma,\sigma)+i(\lambda^a)^T \phi_a\right)~,
\label{eq:varphi}
 \ee
where we denote the pseudo Nambu-Goldstone boson after spontaneous chiral symmetry breaking as $\phi_a$. These dark pions are stable due to flavor symmetry and they serve as good DM candidates.
In the self-consistent mean field approximation one splits up the NJL Lagrangian (\ref{eq:NJL10}) into the sum
$$
\mathcal{L}_{\rm NJL}= \mathcal{L}_{0}+\mathcal{L}_{I}~,
$$
where  $\mathcal{L}_{I}$ is 
normal ordered 
(i.e. $\langle 0\vert \mathcal{L}_{I}\vert 0\rangle =0$), and $\mathcal{L}_{0}$ contains at most fermion bilinears
which are not normal ordered.
After some manipulations, one finds the following form for ${\cal L}_0$
\cite{Holthausen:2013ota}:
\begin{align}
{\cal L}_0 =&
\mbox{Tr}\bar{\psi}\gamma^\mu( i\partial_\mu +g' Q B_\mu)\psi-
\left(\sigma+y S-\frac{G_D}{8G^2}\sigma^2\right)
\mbox{Tr}\bar{\psi}\psi -i \mbox{Tr}\bar{\psi}\gamma_5 \phi\psi -\frac{1}{4G}\sum_{a=1}^8\phi_a\phi_a \nn\\
& -\frac{3\sigma^2}{8G}+\frac{G_D}{8G^2}\left(
- \mbox{Tr}\bar{\psi}\phi^2 \psi+\sum_{a=1}^8\phi_a\phi_a \mbox{Tr}\bar{\psi}\psi +i \sigma\mbox{Tr}\bar{\psi}\gamma_5 \phi \psi+ \frac{\sigma^3}{2G}+
\frac{\sigma}{2G} \sum_{a=1}^8(\phi_a)^2\right).
\label{eq:L0}
\end{align}
Note that this Lagrangian no longer  contains the four and six fermi interactions.
At the non-trivial lowest order only ${\cal L}_0$ is relevant
for the calculation of  the effective potential, the
DM mass $m_{\rm DM}$ and the DM interactions.
The mass spectrum for all the CP-even particles, namely $h,S$ and $\sigma$ can be obtained from the minimum of the effective potential, once the free parameters of the model i.e. $y,\lambda_H,\lambda_{HS},\lambda_S$ 
are given. See Ref.~\cite{Holthausen:2013ota} for more details in the calculation of the effective potential. The dimensionless couplings $y,\lambda_H,\lambda_{HS},\lambda_S$  are required to satisfy perturbativity 
and vacuum stability up the Planck scale. Once the global minimum of the effective potential is obtained, the effective couplings between the bosons and the DM properties are determined.
The $U(1)_Y$ coupling does not contribute to the effective potential and the mass matrix for $h,S,\sigma$ in the lowest order. 
 
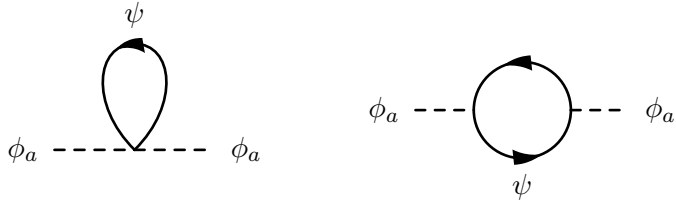
\begin{figure}[t]
\begin{center}
\subfloat{
\begin{fmffile}{dm1}
	        \begin{fmfgraph*}(60,50)
	            \fmfleft{g1}
                    \fmflabel{$\phi_a$}{g1}
	            \fmfright{g2}
	            \fmflabel{$\phi_a$}{g2}
	            \fmf{dashes}{g1,t1}
	            \fmf{dashes}{t1,g2}
		    \fmf{fermion,right=3,label=$\psi$,l.side=right,tension=0.5}{t1,t1}
	        \end{fmfgraph*}
	    \end{fmffile}}
\hspace{2cm}
\subfloat{
\begin{fmffile}{dm2}
	        \begin{fmfgraph*}(80,80)
		    \fmfleft{g1}
                    \fmflabel{$\phi_a$}{g1}
	            \fmfright{g2}
	            \fmflabel{$\phi_a$}{g2}
	            \fmf{dashes}{g1,t1}
	            \fmf{dashes}{t2,g2}
		    \fmf{fermion,label=$\psi$,right=1,l.side=right,tension=0.3}{t1,t2}
		    \fmf{fermion,right=1,l.side=right,tension=0.3}{t2,t1}
	        \end{fmfgraph*}
	    \end{fmffile}}
\end{center}
\caption{One-loop contributions of the heavy dark fermions to the DM mass.}
\label{con1-fig:dmdmgammagamma}
\end{figure}

As we have mentioned that the CP-odd pseudo Nambu-Goldstone bosons $\phi_a$ are the DM candidates for our model, let us investigate their properties in more details. Like the CP-even bosonized $\sigma$ field, the DM 
particles have no tree level kinetic term and their masses are defined as the zero of the inverse propagator
\begin{align}
\Gamma_{\phi}(p^2)=&-\frac{1}{2G}+\frac{G_D \langle \sigma\rangle}{8G^3} +\frac{G_D N_c}{G^2} \int \frac{d^4 k}{i(2\pi)^4}\frac{M}{(k^2-M^2)} \nonumber \\
&+2 N_c\left(1-\frac{G_D \langle \sigma\rangle}{8G^2}\right)^2 \int \frac{d^4 k}{i(2\pi)^4}
\frac{\operatorname{Tr}(\slashed{k}-\slashed{p}+M)
\gamma_5(\slashed{k}+M)\gamma_5}{((k-p)^2-M^2)(k^2-M^2)},
\label{con1-sigma-dm}
\end{align}
where $M=\sigma+yS-G_D \sigma^2/8G^2$ is the constituent hidden sector fermion mass when all the CP-even scalar fields obtained their vacuum expectation values (VEV).
The first two terms in Eq.~\eqref{con1-sigma-dm} stem from the tree level effective Lagrangian \eqref{eq:L0} while the heavy dark fermions contribute to the one-loop radiative correction for the DM inverse propagator.
The relevant one-loop diagrams are given in Fig.~\ref{con1-fig:dmdmgammagamma}.
From Eq.~\eqref{con1-sigma-dm} the DM mass and its wave function renormalization constant
can be calculated
\begin{align}
\Gamma_{\phi}(m_{\mathrm{DM}}^2)=&0~,~
Z_{\phi}^{-1} = \left.\frac{d \Gamma_{\phi}(p^2)}{d p^2}~\right|_{p^2=m_{\mathrm{DM}}^2}.
\label{con1-wave0}
\end{align}
As $y\rightarrow 0$, the chiral symmetry of the fermions should be restored and $m_{\rm DM} \rightarrow 0$ \footnote{In this case $\phi$ is a true Nambu-Goldstone boson.}, hence the size of the DM mass is controlled by
the Yukawa coupling $y$. The additional $U(1)_Y$ coupling however does not contribute to the DM mass.

\section{Relic abundance of DM and its Direct detection}
\begin{figure}
  \includegraphics[width=12cm]{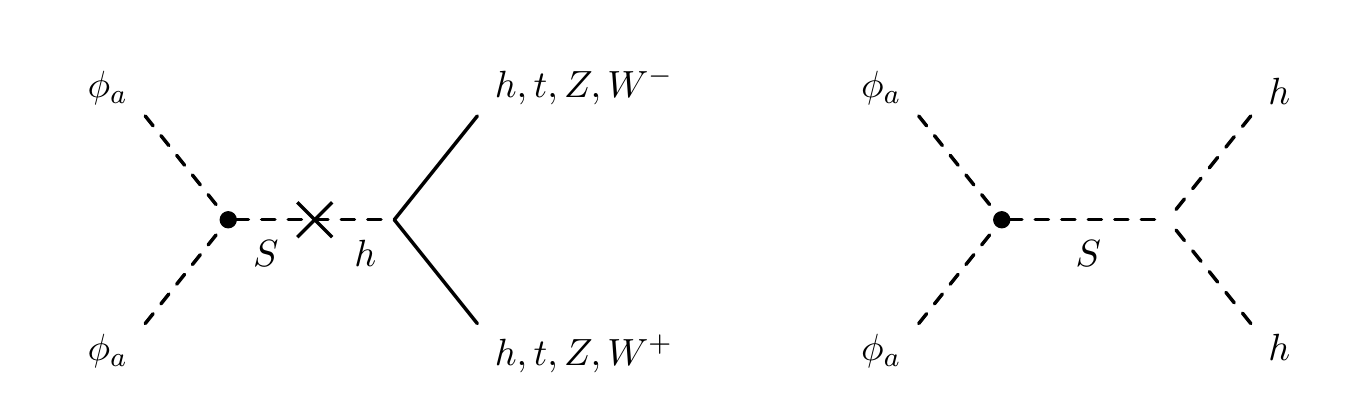}
    \includegraphics[width=4cm]{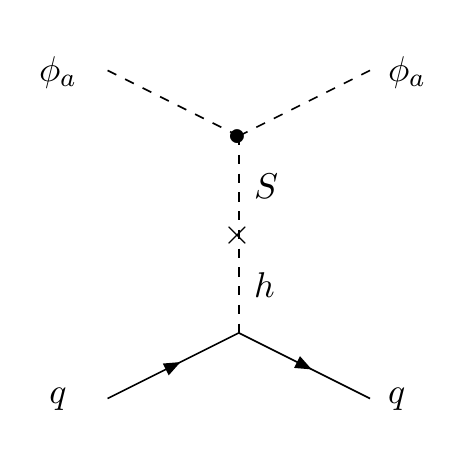}
\caption{\label{dmdmSMSM}\footnotesize
The left and middle diagrams are the s-channel DM annihilation diagrams.
The right diagram  contributes to the DM scattering off the nucleon.
The coupling marked with a dot is a one-loop
three-point vertex given in \cite{Holthausen:2013ota}.
}
\end{figure}
Before we start to compute the relic abundance  $\Omega h^2$, let us  discuss the parameter space. In our previous paper \cite{Holthausen:2013ota} the dimensionless coupling constants,
$y,\lambda_{S,HS,H}$, are constrained by the vacuum stability and by the absence of the Landau pole. It turns out that  with a non-zero $Q$ (at least for $Q\lesssim 1/3$) the allowed parameter
space does not practically change. Note that the annihilation processes of DM occur at the one-loop level through the one-loop $\phi\mbox{-}\phi\mbox{-}S$ amplitude and the one-loop $\phi\mbox{-}\phi\mbox{-}S\mbox{-}S$ amplitude
(if $m_S < m_{\rm DM}$). 
The $\phi\mbox{-}\phi\mbox{-}S$ amplitude can be calculated from the one-loop diagram
\begin{align}
\Gamma_{\phi\phi S}=& 4N_c y\left(1-\frac{G_D \langle\sigma\rangle}{8G^2}\right)^2 \int \frac{d^4 k}{i(2\pi)^4}\frac{\operatorname{Tr}(\slashed{k}+M)\gamma_5(\slashed{k}-\slashed{p}+M)
(\slashed{k}+\slashed{p'}+M)\gamma_5}{((k-p)^2-M^2)(k^2-M^2)((k+p')^2-M^2)} \nn \\
&+N_c y\frac{G_D}{4G^2} \int \frac{d^4 k}{i(2\pi)^4}\frac{\operatorname{Tr}(\slashed{k}-\slashed{p'}+M)
(\slashed{k}+\slashed{p}+M)}{((k-p')^2-M^2)((k+p)^2-M^2)},
\label{con1-dm-dm-s}
\end{align}
which is crucial for determining the relic abundance and the direct detection cross section of the DM. The momenta $p,p^\prime$ represent the incoming momenta of the dark pions. 
The one-loop effective couplings are represented as $\bullet$ in Fig.~\ref{dmdmSMSM}. 
These amplitudes are small for small $y$ as the amplitudes scale like $\mathcal{A}(\phi\phi\rightarrow S)\sim y$ and $\mathcal{A}(\phi\phi\rightarrow SS)\sim y^2$ respectively. 
As mentioned above, the size of the DM mass is controlled by $y$, i.e. a small $y$ implies a small DM mass and for a larger $y \gtrsim 0.2$ the DM mass $m_{\rm DM}$ 
can become larger than the fermion constituent mass $M$, which will develop imaginary parts in these one-loop amplitudes. In \cite{Holthausen:2013ota} we forbad the 
occurrence of the imaginary parts, yielding an upper bound on $y$ for a given set of $\lambda_{S,HS,S}$. 
As the parameter $y$ is bounded from above, the $\phi\mbox{-}\phi\mbox{-}S\mbox{-}S$ amplitude contributes negligibly to the relic abundance calculation and only the $\phi\mbox{-}\phi\mbox{-}S$ amplitudes
are important.
However in this parameter space we have found that the only way to enhance the annihilation rate of DM is via a resonance effect in the s-channel annihilation processes shown in Fig.~\ref{dmdmSMSM} (left and middle).
That is, $2m_{\rm DM} \simeq m_S$ has to be satisfied. The direct detection rate of DM is however strongly suppressed ($\lesssim \unit[10^{-48}]{cm^2}$) 
because it is a t-channel process shown in Fig.~\ref{dmdmSMSM} (right), constraining the parameter space into phenomenologically unattractive corner.

\begin{figure}
        \includegraphics[width=10cm]{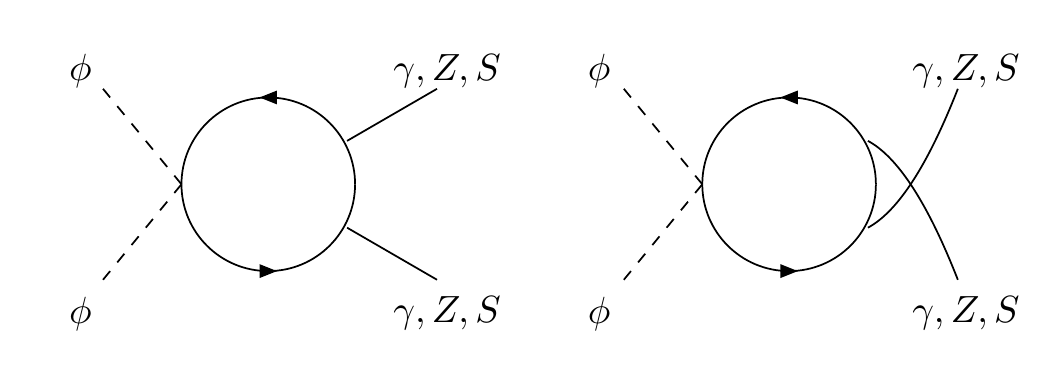}
        \caption{\label{dmdmgg5}\footnotesize
The DM annihilation with $\phi\phi\bar{\psi}\psi$
coupling (a)}
  \includegraphics[width=10cm]{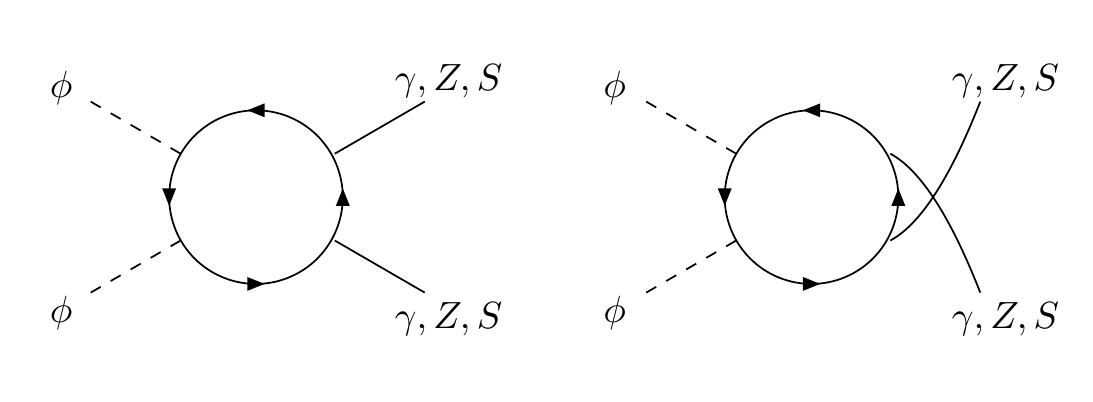}
    \includegraphics[width=10cm]{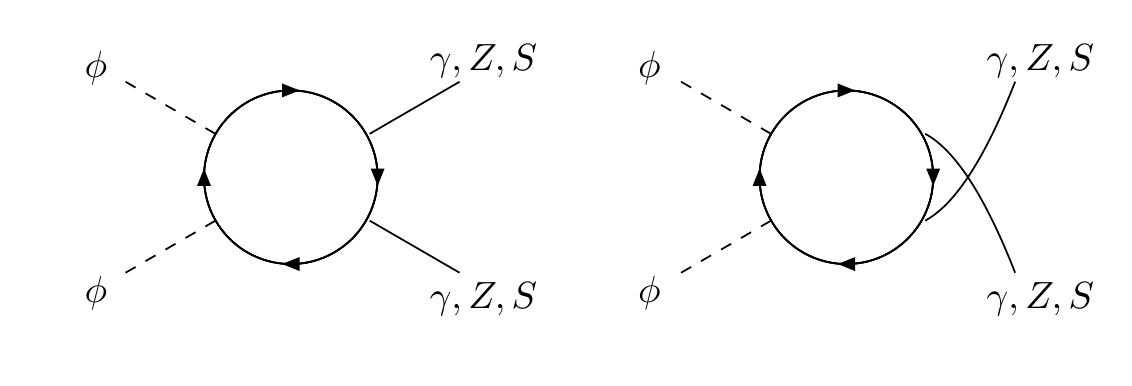}
            \caption{\label{dmdmgg1}\footnotesize
The DM annihilation (b)}
        \includegraphics[width=10cm]{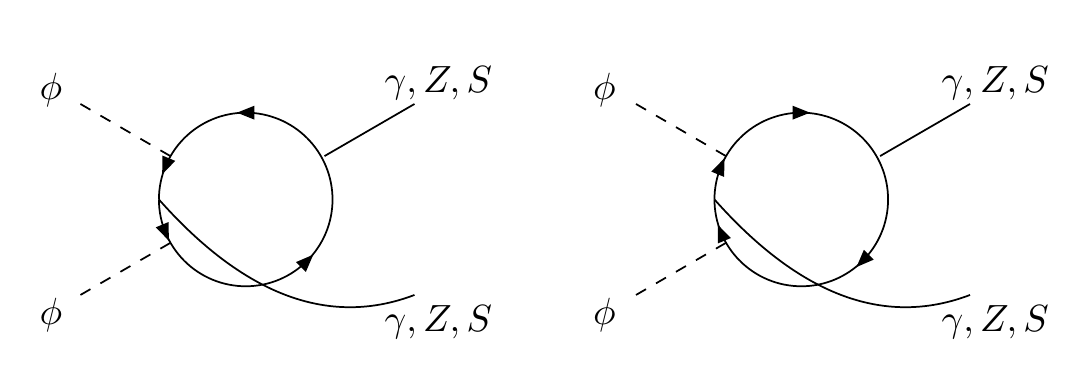}
           \caption{\label{dmdmgg3}\footnotesize
The DM annihilation (c).}
\end{figure}

\begin{figure}
  \includegraphics[width=7cm]{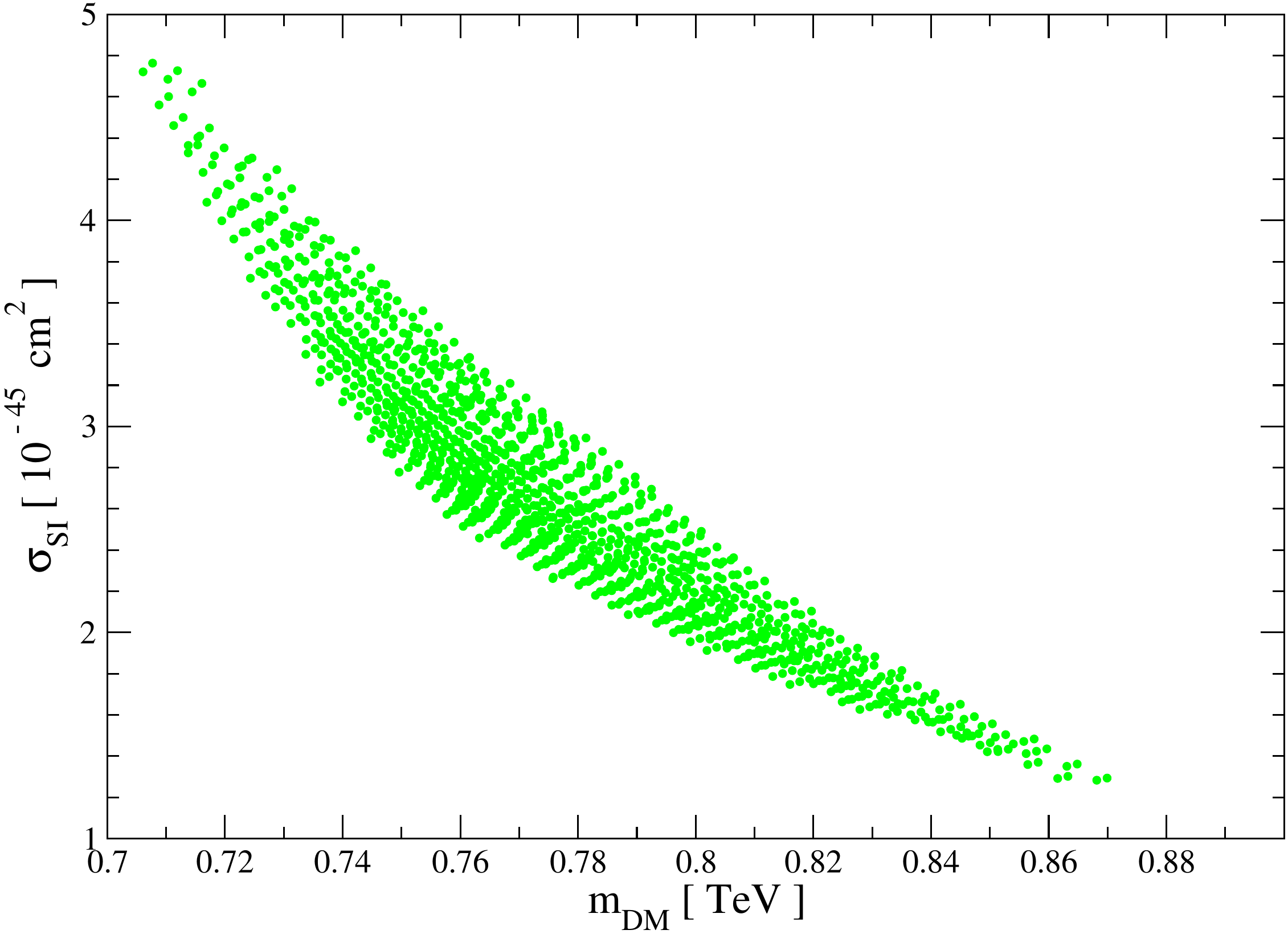}
   \includegraphics[width=7cm]{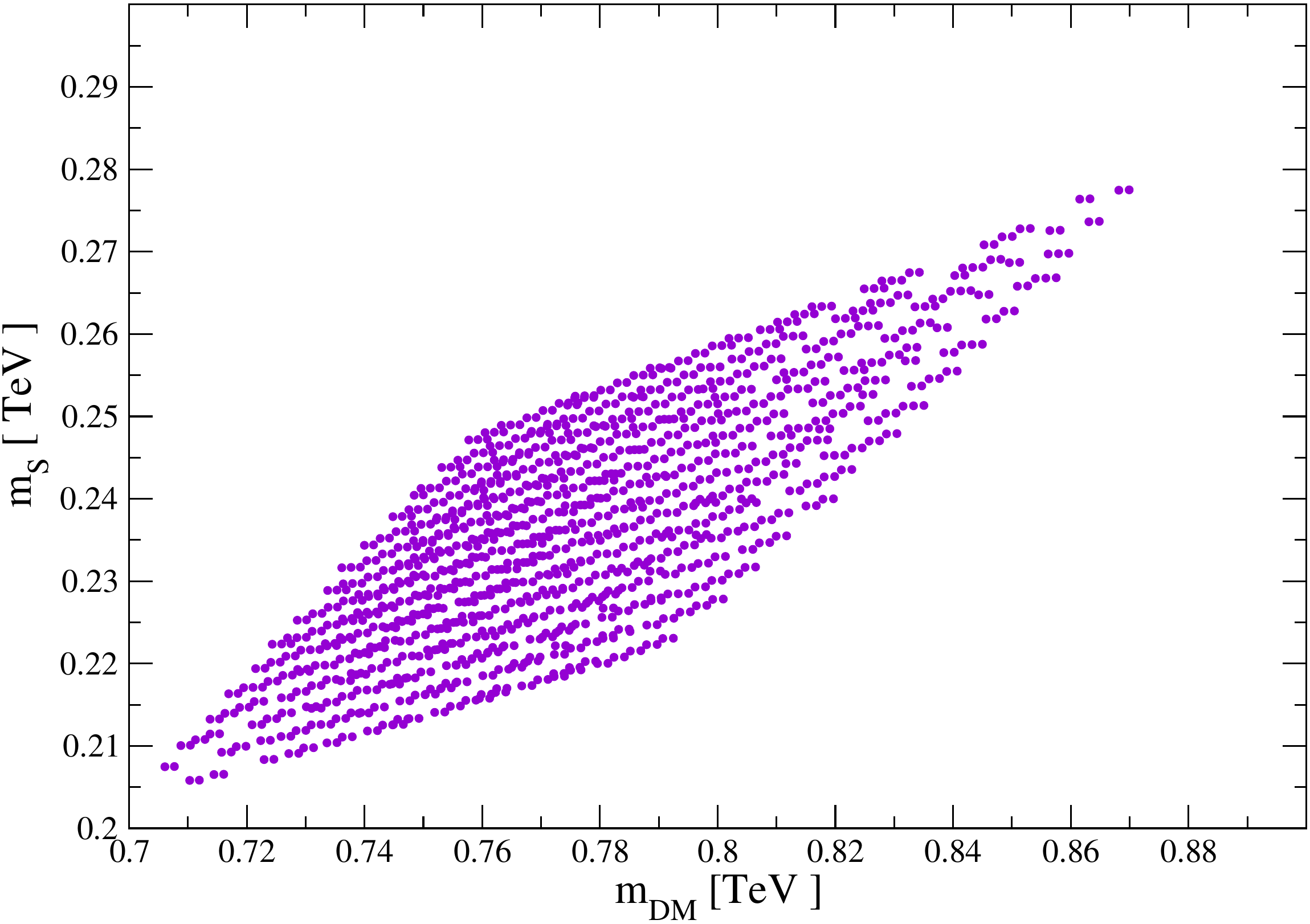} 
  \caption{\label{mdm-sigma}\footnotesize
Left:The spin-independent cross section off nucleon
against $m_{\rm DM}$,
where $\Omega h^2=0.1187
\pm 0.005(3\sigma)$ \cite{Ade:2013ktc} is imposed. The XENON100 \cite{Aprile:2012nq} and LUX \cite{Aprile:2012zx} limits are $\sim \unit[10^{-44}]{cm^2}$ for $m_{\rm DM}
=\unit[0.7]{TeV}$,
while the XENON1T sensitivity is two orders of magnitudes
higher than that of XENON100 \cite{Aprile:2012zx}.
Right: The mass of the singlet $S$ against $m_{\rm DM}$.
If  $m_{\rm DM} < m_S$, then $\sigma_{\rm SI}
\lesssim \unit[10^{-48}]{cm^2}$ \cite{Holthausen:2013ota} .}
\end{figure}

In this paper we allow the occurrence of the imaginary parts in the one-loop diagrams, as they are related to the real parts due to the dispersion relation, which has proven to be
successful in describing the QCD hadron physics (see \cite{Hatsuda:1994pi} for instance). We set the upper bound at $m_{\rm DM} < 2 M$, which should be compared with 
$m_{\eta'}=0.958$ GeV and $M_s =0.5$ GeV in the usual QCD physics, where $M_s$  is the constituent mass of the strange quark. In fact, in the optimistic range $y\gtrsim 0.4$ with   
$m_S < m_{\rm DM}$, the $\phi\mbox{-}\phi\mbox{-}S\mbox{-}S$ amplitude (which is generated at one-loop as shown in Fig.~\ref{dmdmgg5} $\sim$ \ref{dmdmgg3}) is no longer
small  and can become large enough to give a correct relic abundance of DM. Therefore, we choose below the parameter space $y\gtrsim 0.4$ and open the channel $\phi\phi\to SS$.
In this parameter region, the s-channel processes contributed by $\phi\mbox{-}\phi\mbox{-}S$ amplitudes are negligibly suppressed and can be ignored.
The annihilation diagrams in Fig.~\ref{dmdmgg5} $\sim$ \ref{dmdmgg3} for $\phi\phi\to SS$ yield
\begin{align}
\mathcal{A}(\phi\phi \to SS) = & 2 N_c y^2 \left[ \frac{G_D}{4G^2}I_a + \left( 1-\frac{G_D \langle \sigma \rangle}{8G^2} \right)^2 \left( 4I_b+2I_c \right) \right],
\end{align}
where $I_i$ represents the integral for the respective $i$th loop diagram in Fig.~\ref{dmdmgg5} $\sim$ \ref{dmdmgg3}.
We obtain the DM annihilation cross section 
\begin{align}
\langle v\sigma(\phi \phi \to SS)\rangle &=
\frac{Z_\phi^2}{32 \pi m_{\rm DM}^3}\vert {\cal A}(\phi \phi \to SS)\vert^2 \left(1-\frac{m_S^2}{m_{\rm DM}^2} \right)^{1/2}~,
\end{align}
where $Z_\phi$ is given in Eq.~\eqref{con1-wave0}.
We do not include the annihilation modes into $\gamma\gamma,\gamma Z,ZZ$ as the annihilation cross section of these modes is proportional to $\alpha^2 Q^4$ (see Eq.~\eqref{eq:gammaann}) while the DM annihilation cross section to $S$ particles is dominated by $y^4$. Unless the electric charge $Q\gtrsim1$, the annihilation modes into $\gamma\gamma,\gamma Z,ZZ$ can be ignored in the relic abundance calculation (see also the comment in the footnote on page 12). The annihilations into these modes are calculated in the next section.
We find, imposing the constraint on the relic
abundance $\Omega h^2 =0.1187
\pm 0.005(3\sigma)$ \cite{Ade:2013ktc},
 that the spin independent annihilation cross section
is just below the XENON100 \cite{Aprile:2012nq}
and LUX \cite{Akerib:2013tjd}  constraints and above the
 XENON1T sensitivity \cite{Aprile:2012zx}. This is shown in Fig.~\ref{mdm-sigma}. 

The DM mass $m_{\rm DM}$ is constrained in the present model.
The lower limit  $m_{\rm DM} \gtrsim \unit[0.7]{TeV}$ comes from the fact that
$y$ has to be large
enough so that the size of the annihilation process $\phi\phi\to SS$
yields a correct relic abundance of DM.
If on the other hand $y$ is too large, 
the annihilation cross section 
into two singlet scalars becomes too large so that
the relic abundance falls below the observed value.
The Yukawa coupling $y$ is also constrained from above 
to avoid the triviality bound,
which gives the upper limit
$m_{\rm DM}\lesssim \unit[0.9]{TeV}$.

Note that because of the $SU(3)_V$ flavor symmetry and 
an accidental $U(1)_{\rm hB}$  (hidden baryon number),
not only the DM candidates, but also the lightest hidden baryons 
are stable. 
In the case that $N_c=N_f=3$ in the hidden sector and $Q=1/3$ for the hidden fermions,
there is no stable hidden hadron with a fractional electric charge.
The hidden mesons for our model are neutral, while the charge of the hidden baryons formed by three hidden fermions is one if $Q=1/3$.
There might be a tiny amount of relic stable hidden baryons and anti-baryons in the universe, which if a large number of them are not annihilated,
could spoil the large scale structure formation. 
Let us roughly estimate fraction of this hidden baryon. As
the hidden sector is  described by a scaled-up QCD, so that
because the coupling  $G_{\phi B \bar{B}}$ is dimensionless,
the hidden meson-baryon coupling $G_{\phi B \bar{B}}$ is approximately the 
same as in QCD, i.e.  $G_{\phi B \bar{B}}\sim 13$. Using this fact, 
we have estimated the relic abundance $\Omega_{hB}h^2$ of the
hidden baryons to be $\sim 10^{-4}$ for the hidden baryon mass
of $\unit[3]{TeV}$. Note that this result is independent of Q. Therefore, we may fairly ignore the stable charged hidden baryons
in discussing the relic abundance of DM.

\section{Restoring gauge invariance }
The cutoff $\Lambda$ breaks gauge invariance explicitly and to restore gauge invariance we have to subtract non-gauge invariant terms
from the original amplitude.
In  renormalizable theories there is no problem to define
a finite renormalized gauge invariant amplitude, i.e.
In  the limit of $\Lambda\to \infty$
the gauge non-invariant terms are a finite number of local 
terms,
 which can be cancelled by the corresponding
local counter terms so that the subtracted amplitude
is, up to its normalization,  independent of the regularization scheme
(see for instance \cite{Jauch}).
To achieve such a uniqueness
in  cutoff theories, one needs 
an additional prescription.
For instance, we can define the real part of an amplitude using
dispersion relation, as it was  done  in the original paper 
by Nambu and Jona-Lasinio \cite{Nambu:1961tp}
(see also \cite{Hatsuda:1994pi}).
This procedure yields a gauge invariant real part of the amplitude 
in one-loop,
because the imaginary part of the amplitude is gauge invariant
in  one-loop order.
In this method, however, the one-loop tadpole diagram cannot be reproduced from its imaginary part as the tadpole diagram does not contain any imaginary part. 
Another way\footnote{ $\zeta$-function regularization was also  used  in 
\cite{Suganuma:1990nn}
to obtain  a gauge invariant effective potential 
in the presence of the electromagnetic field as an external field.}  is to utilize a gauge invariant regularization
such as the Pauli-Villars regularization \cite{Klevansky:1991ey,Klevansky:1992qe} which preserves gauge invariance by construction but breaks chiral symmetry explicitly. The drawback is however, 
for a finite regulator mass,
it is not clear whether the breaking of chiral symmetry 
results from  the regulator or from  non-perturbative effect.
Moreover, the regulator fields are ``ghost'' fields, which 
are not completely decoupled at a finite cutoff $\Lambda$.

We will propose another method,
 which we call ``\emph{least subtraction procedure}''. In the NJL model as a cutoff theory 
the cutoff $\Lambda$ is a physical parameter, and a finite $\Lambda$ is 
essential
to describe effectively D$\chi$SB.  If we subtract too much from the amplitude
to restore gauge invariance, we may lose information on 
non-perturbative effects.
Therefore, we stress that
we keep the subtraction terms to the minimum as necessary.
The details of least subtraction procedure is given in Appendix \ref{sec:app},
where we consider the  photon self-energy, 
the $S\mbox{-}\gamma\mbox{-}\gamma$ as well as the $\phi\mbox{-}\phi\mbox{-}\gamma\mbox{-}\gamma$
vertex functions. The results are applied to the next section
for the calculation of the DM annihilation cross section
into two  $\gamma$'s.

\section{Monochromatic \mbox{\texorpdfstring{$\gamma$}{gamma}}-ray line from DM annihilation}
The charge $Q$ of the hidden fermion  is a free parameter.
It can be constrained from the indirect detection
of DM, e.g. the upper bound 
on $\sigma(\phi\phi\to \gamma\gamma)$ for 
$\gamma$-ray lines given in 
\cite{Ackermann:2012qk,Gustafsson:2013fca,Abramowski:2013ax}.
The four-point $\phi\mbox{-}\phi\mbox{-}\gamma\mbox{-}\gamma $ coupling\footnote{The  $U(1)_Y$ 
gauge invariance and  the $SU(3)_F$ flavor symmetry together  with 
the reality
of $\phi$ forbid the existence of the  $\phi\phi B_\mu$ coupling.}
 is generated 
at one-loop as is 
shown in Fig.~\ref{dmdmgg5} $\sim$ \ref{dmdmgg3},
which predicts the DM annihilation into two monochromatic photons
of energy $ m_{\rm DM}$.
Similar processes have been calculated  in a universal extra dimension
model \cite{Bertone:2009cb}, for instance.
In Appendix \ref{sec:app} it is  shown how to
restore gauge invariance of 
the  four-point  amplitude, with the result given in \eqref{AmunuR}.
If we neglect the mass of $Z$ against $m_{\rm DM}$,
 the four-point   functions 
${\cal A}^R_{\mu\nu}(\gamma Z)$ and ${\cal A}^R_{\mu\nu}(Z Z)$ can
be approximated 
by (\ref{AmunuR})  as well,  with the replacement of
$e^2$ by $-e^2 t_W$ and $e^2 t_W^2$, respectively.

For $p=p'=(m_{\rm DM},\bf 0)$ the photon momenta take the form
$k=(m_{\rm DM},\bf k)$ and $k'=(m_{\rm DM},-\bf k)$, with
their polarization tensors
 $\epsilon(k)=(0,\mbox{\boldmath  $\epsilon$}(k))$ and 
$\epsilon(k')=(0, \mbox{\boldmath  $\epsilon$}(k'))$ satisfying
 \be
0&=& \epsilon(k)\cdot k=
\epsilon(k)\cdot k'=
\epsilon(k)\cdot p=
\epsilon(k)\cdot p'\\
0&=&
\epsilon(k')\cdot k=
\epsilon(k')\cdot k'=
\epsilon(k')\cdot p'=\epsilon(k')\cdot p'~,
\ee
respectively.
Therefore, only $g_{\mu\nu}$ terms of the subtracted
gauge invariant four-point function $\Gamma_{\mu\nu}(\gamma\gamma)$ contributes:
\be
\Gamma_{\mu\nu}(\gamma\gamma)
&=& g_{\mu\nu}\left({\cal A}^{R(a)}+{\cal A}^{R}_g\right)
=i\frac{\alpha  }{ \pi}Q^2 ~ g_{\mu\nu}{\cal A}(\gamma\gamma)~,
\ee
where ${\cal A}^{R(a)}$ (defined in (\ref{ARa})) is the contribution 
from Fig.~\ref{dmdmgg5}, while
${\cal A}^{R}_g$ (defined in (\ref{ARg}))
is the contribution from Fig.~\ref{dmdmgg1} and \ref{dmdmgg3}.
The other ones can be approximated as
\be
\Gamma_{\mu\nu}(a~b)&\simeq & 
i\frac{\alpha  }{ \pi}Q^2 ~ g_{\mu\nu}{\cal A}(\gamma\gamma)
\times \left\{\begin{array}{cc} & a~b\\
-t_W&\gamma ~Z  \\
t_W^2&Z ~Z
 \end{array}~.
\right.
\ee

\begin{figure}
  \includegraphics[width=8cm]{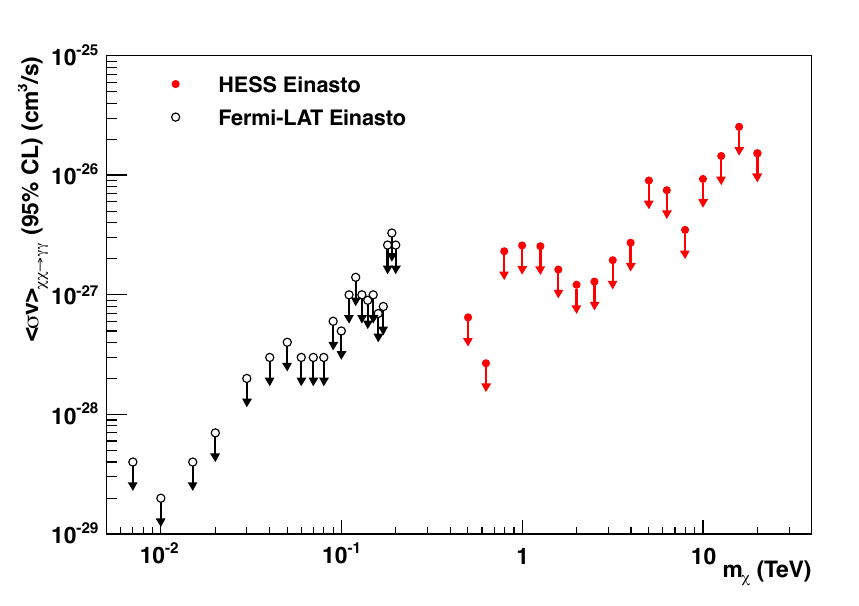}
 \includegraphics[width=7.5cm]{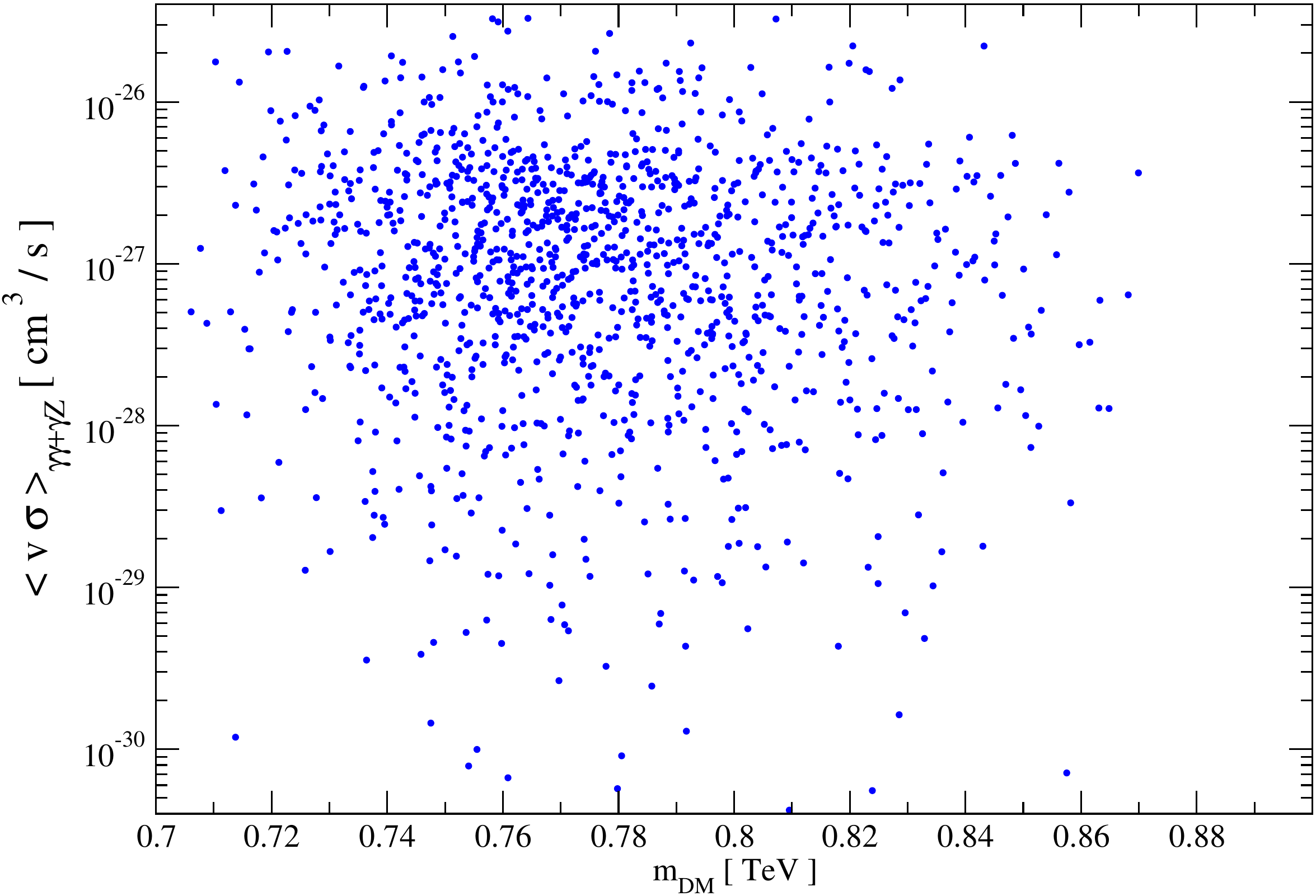}
\caption{\label{dmdmgg}\footnotesize
Left: The Fermi Lat \cite{Gustafsson:2013fca} (black)
and HESS\cite{Ackermann:2012qk} (red) 
upper bounds  on the velocity-averaged  DM annihilation
cross section for monochromatic $\gamma$-ray lines, where this graph is taken from  \cite{Ackermann:2012qk}.
Right: the velocity-averaged  DM annihilation cross section
$\langle v\sigma\rangle_{\gamma\gamma+\gamma Z}$
as a function of $m_{\rm DM}$ with $Q=1/3$, where $\Omega h^2 =0.1187
\pm 0.005(3\sigma)$ \cite{Ade:2013ktc} is imposed. Since $\langle v\sigma\rangle_{\gamma\gamma+\gamma Z}$ is proportional to
$Q^4$,  our calculations can be simply 
extended to the case of an arbitrary $Q$.}
\end{figure}
Then  (the s-wave part of) the corresponding   velocity-averaged  
annihilation cross sections are given by
\be
\langle v\sigma(\phi\phi\to a~b)\rangle &= & 
\frac{\alpha^2 Q^4  Z_\phi^2}{16 \pi^3 m_{\rm DM}^2}
{\cal A}^2(\gamma\gamma)\times \left\{\begin{array}{cc} & a~b\\
(1/2)& \gamma~\gamma \\
t_W^2(1-m_Z^2/4 m_{\rm DM}^2) &\gamma ~Z \\
(3/4)t_W^4(1-m_Z^2/m_{\rm DM}^2)^{1/2} & Z~Z
 \end{array}~,
\right.
\label{eq:gammaann}
\ee
where $Z_\phi$ is the wave function renormalization constant
which is given in \cite{Holthausen:2013ota}.
The  energy $E_\gamma$ of $\gamma$-ray line produced 
in the annihilation into $\gamma Z$
is $m_{\rm DM}(1-m_Z^2/4 m_{\rm DM}^2)$.
In practice, however, due to finite detector energy resolution
this line cannot be distinguished 
from the $E_\gamma=m_{\rm DM}$ line.
Therefore, we simply add both cross sections.
So we compute
$\langle v\sigma\rangle_{\gamma\gamma+\gamma Z}
=\langle v\sigma(\phi\phi\to \gamma\gamma)\rangle+
\langle v\sigma(\phi\phi\to \gamma Z )\rangle$
with $Q=1/3$
as  a function  of $m_{\rm DM}$ for different values of $\lambda_H,~
\lambda_S$ and $\lambda_{HS}$.
As noticed in the previous section, we have not included
the annihilation modes into $\gamma\gamma, \gamma Z, ZZ$
in calculating the relic abundance.
In this way we can obtain a separate information on the size of the 
annihilation cross section producing  the
line $\gamma$-ray spectrum
of DM in this model
\footnote{
The contribution can become important for$ 
\langle v\sigma\rangle_{\gamma\gamma+\gamma Z}
\gtrsim   \langle v\sigma(\phi \phi \to SS)\rangle \simeq 8\times \unit[10^{-27}]{cm^3/s}$.
But this approximate inequality can not be satisfied for $m_{\rm DM}$
between $0.7$ TeV and $0.9$ TeV, if  the HESS constraint for 
$m_{\rm DM}=0.8$ TeV, i.e.
$\langle v\sigma\rangle_{\gamma\gamma+\gamma Z}
\lesssim 2\times 10^{-27}~\mbox{cm}^3/\mbox{s}$,
is satisfied for this range of $m_{\rm DM}$.
If the HESS constraint for 
$m_{\rm DM}=0.8$ TeV does not apply and there is no cosmological constraint
for this range of $m_{\rm DM}$,
we should  control the size of 
$\langle v\sigma\rangle_{\gamma\gamma+\gamma Z}$ by varying $Q$
when the approximate inequality above  is satisfied.
}.

As we see from Fig.~\ref{dmdmgg} (left) strong constraints are given for $m_{\rm DM}\simeq
0.6\, \unit[(0.5)]{TeV}$:
$\langle v\sigma\rangle_{\gamma\gamma+\gamma Z}
\lesssim 3\, (7)\times \unit[10^{-28}]{cm^3/s}$.
Since our DM is heavier than $\unit[0.7]{TeV}$ (see Fig.~\ref{mdm-sigma}),
these strong constraints do not apply.
Above $\unit[0.7]{TeV}$, the upper  bound is about one order
of magnitude larger than that for  $m_{\rm DM}=0.6$
TeV, so that the constraints 
can well be satisfied even for $Q >1/3$,
as we can see from Fig.~\ref{dmdmgg} (right).
An interesting feature of the present model is that 
the $\gamma$-ray line energy is constrained 
between $\sim \unit[0.7]{TeV}$ and $\sim \unit[0.9]{TeV}$,
because the DM mass $m_{\rm DM}$ is constrained as it is explained in the
previous section.
 Another feature of the model related to
  $\gamma$-ray lines is that the production cross
  section of $\gamma$-ray lines is in the same order in $1/N$ expansion 
  (i.e. in one-loop order) as the 
  total annihilation cross section of DM.
    That is, 
  $\langle v\sigma\rangle_{HH, f\bar{f},WW,\cdots}
\sim \langle v\sigma\rangle_{\gamma\gamma+\gamma Z}$
in the present model.
This is similar to one of three exceptions,
forbidden channels, considered in \cite{Tulin:2012uq}.
In the case of  the forbidden channels
the tree-level processes are kinematically forbidden,
which should be contrasted to the
present case in which
the Nambu-Goldstone DM has no contact with the messenger field $S$
at the tree-level.

The differential $\gamma$-ray flux is given by
\be
\frac{d\Phi}{d E_\gamma} &\propto &
 \langle v\sigma\rangle_{\gamma\gamma}
\frac{d N^{\gamma \gamma}}{d E_\gamma} +
\langle v\sigma\rangle_{\gamma Z}
\frac{d N^{\gamma z}}{d E_{\gamma Z}} 
\simeq \langle v\sigma\rangle_{\gamma\gamma+\gamma Z}
~\delta(E_\gamma-m_{\rm DM})~.
\ee
Prospects observing such line spectrum is discussed in detail in
\cite{Bertone:2009cb,Laha:2012fg}. Obviously,   with an increasing energy resolution 
the  chance for the observation increases.
Observations of  $\gamma$-ray lines of energies
 between $\sim \unit[0.7]{TeV}$ and $\sim\unit[0.9]{TeV}$ TeV 
 not only fix the charge of the hidden sector fermion, but also yields
 a first experimental hint on the hidden sector.

\section{Conclusion}

The Nambu-Goldstone theorem predicts in the presented model 
for the hidden sector, where chiral symmetry is dynamically broken 
and hence a scale is created, the existence of a DM 
candidate.
This generated scale is transmitted to the SM sector via a real SM singlet 
scalar $S$ to trigger spontaneous breaking of electroweak gauge symmetry.
With a non-zero $U(1)_Y$ hypercharge $Q$ of the hidden sector
fermion the hidden sector is no longer dark, and new possibilities 
to test experimentally the hidden sector are open.
We studied in this paper the possibility of DM annihilation and
found that this model allows DM to annihilate into two photons, 
producing a $\gamma$-ray line spectrum.
We found that the $\gamma$-ray line energy must be  between 
$\unit[0.7]{TeV}$ and $\unit[0.9]{TeV}$ with  the velocity-averaged 
annihilation cross section $\unit[10^{-30} \sim 10^{-26}]{cm^3/s}$ 
for $Q=1/3$, which satisfies easily the recent limits given by 
Fermi LAT \cite{Ackermann:2012qk,Gustafsson:2013fca} and 
HESS \cite{Abramowski:2013ax}.

With a non-zero $Q$ the hidden sector is doubly connected with the 
SM sector. The connection via photon and $Z$ opens possibilities to 
probe the hidden sector at collider experiments such as $e^+ e^-$ 
collision \cite{Kanemura:2011nm}. In the parameter range, where the 
annihilation of DM into two singlets $SS$ is dominant and a 
correct relic abundance of DM is obtained, the constituent mass $M$ 
of the fermion is comparable with $m_{\rm DM}$, i.e.
$\unit[0.7]{TeV}\lesssim M\lesssim \unit[0.9]{TeV}$. This is the energy 
region of hidden hadron physics and the scale of the hidden sector itself is 
$\sim \unit[0.7]{TeV}$, compared to $\Lambda_{\rm QCD}\approx \unit[1]{GeV}$.
The hidden strong interaction becomes therefore perturbative at about 
one order of magnitude above this energy region, $\gtrsim \unit[10]{TeV}$,
and the hidden fermion becomes massless and could be produced directly 
to yield hidden sector jets at collider experiments.

\vspace{0.5cm}
\noindent {\bf Acknowledgements:} We would like to thank 
Teiji Kunihiro for useful discussions.
J.~K. would like to thank the theory group of the 
Max-Planck-Institut f\"ur Kernphysik in Heidelberg for their hospitality.
J.~K. is partially supported by the Grant-in-Aid for Scientific Research (C) from the Japan Society for Promotion of Science (Grant No.22540271).
K.~S.~L. acknowledges support by the International Max Planck Research School for Precision Tests of Fundamental Symmetries.

\appendix

\section{Least Subtraction Procedure \label{sec:app}}
Here we elucidate least subtraction procedure
which can be applied to any 
cutoff theory in principle  to obtain gauge invariant amplitudes.
The basic idea is
to keep the subtraction terms to the minimum necessary.
This works as follows.
Consider an unsubtracted amplitude 
\begin{align}
{\cal A}_{\mu_1\dots\mu_{n_g}}(\Lambda; 
k_1\dots k_{n_g},p_1\dots p_{n_s}), \displaybreak[0]
\label{Amunu}
\end{align}
 with $n_g$ photons and $n_s$ scalars (scalars and axial scalars)
 \footnote{
We impose that the on-shell conditions 
(except for the self-energy) and  the momentum 
conservation for the external momenta are 
satisfied.}. Expand the amplitude in the external momenta
$k$'s and $p$'s:
\begin{align}
{\cal A}_{\mu_1\dots\mu_{n_g}}&=\sum_{m=0}{\cal A}^{(m)}_{\mu_1\dots\mu_{n_g}}~, \displaybreak[0]
\end{align}
where ${\cal A}^{(m)}_{\mu_1\dots\mu_{n_g}}$
consists of  $m$-th order  monomials  of 
 the external momenta.
In general, 
${\cal A}^{(0)}_{\mu_1\dots\mu_{n_g}}
={\cal A}_{\mu_1\dots\mu_{n_g}}(\Lambda; 
0,\cdots,0)$
is non-vanishing and we can subtract it because it is not gauge invariant.
We keep the tensor structure
of ${\cal A}^{(0)}_{\mu_1\dots\mu_{n_g}}$
as the tensor structure of the counter terms for
${\cal A}^{(m)}_{\mu_1\dots\mu_{n_g}} ~(m>0)$
until a new tensor structure for the counter terms is required. We continue this until  no more new tensor structure  is needed.
At each step we stress the minimal number of the  new tensor structures 
for  the counter terms.

 \vspace{0.5cm}
\noindent
$\bullet$\underline{Photon self-energy}\\
As an example  we consider the one-loop photon self-energy.
Using the usual technique, introducing a Feynman parameter $x$
for the denominator of the propagators, going to the Euclidean
momentum space, shifting the internal momentum
appropriately, we obtain
the unsubtracted self-energy tensor
\be
\Pi_{\mu\nu}(\Lambda;k)
&=&i\frac{e^2 Q^2 N_c N_f}{8\pi^2}\int_0^1 dx~\left[~
\frac{4\Lambda^2(1-x)x (g_{\mu\nu}k^2-k_\mu k_\nu)
-\Lambda^4 g_{\mu\nu}}{\Lambda^2+B^2}\right.\nn\\
& &\left.-4(g_{\mu\nu} ~k^2-k_\mu k_\nu)(1-x)x~\ln\left( 1+\Lambda^2/B^2 ) \right)\right]
\label{self0}
\ee
with $B^2=M^2-(1-x)x ~k^2$.
According to least subtraction procedure, we expand
$\Pi_{\mu\nu}(\Lambda;k)$ in $k$ and find
\be
\Pi_{\mu\nu}(\Lambda;k)
&=&i\frac{e^2 Q^2 N_c N_f}{8\pi^2}\left[g_{\mu\nu}{\cal A}_g(\Lambda;k^2)+
k_\mu k_\nu {\cal A}_{kk}(\Lambda;k^2)\right]~,\nn\\
{\cal G}(\Lambda;k^2)&=&{\cal A}_{g}(\Lambda;k^2)+k^2{\cal A}_{kk}(\Lambda;k^2)=
-\frac{\Lambda^4}
{\Lambda^2+M^2}
-\frac{ k^2\Lambda^4}
{6(\Lambda^2+M^2)^2}
-\frac { k^4\Lambda^4}{30(\Lambda^2+M^2)^3}~\nn\\
& & -\frac {k^6\Lambda^4 }{140(\Lambda^2+M^2)^4}
-\frac {k^8 
\Lambda^4}{630(\Lambda^2+M^2)^5}+\cdots~,
\label{calG}
\ee
which would  vanish if the amplitude were  gauge invariant.
Further,
\be
\Pi^{(0)}_{\mu\nu} &=&\Pi_{\mu\nu}(\Lambda;0)
=i\frac{e^2Q^2 N_c N_f}{8\pi^2} g_{\mu\nu}{\cal A}_g(\Lambda;0)=
-i\frac{e^2 Q^2N_c N_f}{8\pi^2} g_{\mu\nu}
\Lambda^4/(\Lambda^2+M^2)~.
\ee
This defines the tensor structure for the counter terms,
because this term is not gauge invariant and has to be subtracted.
Therefore, the subtracted amplitude is ${\cal A}_{g}^R(\Lambda;k^2)
={\cal A}_{g}(\Lambda;k^2)-{\cal G}(\Lambda;k^2)$.
Obviously, in this example, all the non-gauge invariant terms
can be canceled by the  counter terms of this tensor structure.
That is, no more new tensor structure is needed
for the counter terms.

Since  in this example we know the closed expression for the amplitude,
it is not necessary to implement least subtraction procedure.
As we see from (\ref{self0}), the $\Lambda^4 g_{\mu\nu}$ term
is not gauge invariant.
This non-gauge invariant  term, which is a  photon mass function
$\Pi_{\mu\nu}^\Lambda(\Lambda;k)$, can not be made 
gauge invariant
by adding  $k_\mu k_\nu$ terms without introducing
a singularity in $k^2$. Therefore, we have to subtract $\Pi_{\mu\nu}^\Lambda(\Lambda;k)$ from
the self-energy $\Pi_{\mu\nu}(\Lambda;k)$,
in accord with least subtraction procedure as described above.

The  gauge invariant term proportional $\propto\Lambda^2$
in (\ref{self0})
gives a wrong normalization so that further counter terms are needed.
Finally, we have the normalized subtracted
 gauge invariant  self-energy of the photon:
\be
\Pi_{\mu\nu}^R(\Lambda;k)
&=&i\frac{e^2 Q^2 N_c N_f}{8\pi^2}(g_{\mu\nu}k^2-k_\mu k_\nu)\int_0^1 dx~\left[~
\left(\frac{4\Lambda^2(1-x)x }{\Lambda^2+B^2}
-\frac{2\Lambda^2/3}{\Lambda^2+M^2}\right)\right.\nn\\
& &\left.+4(1-x)x~\left\{
\ln\left(1- \frac{(1-x)x k^2}{M^2} ) \right)
-\ln\left( 1-\frac{(1-x)x  k^2}{\Lambda^2+M^2} ) \right)
\right\}
\right]~.
\label{self1}
\ee

\begin{figure}
  \includegraphics[width=10cm]{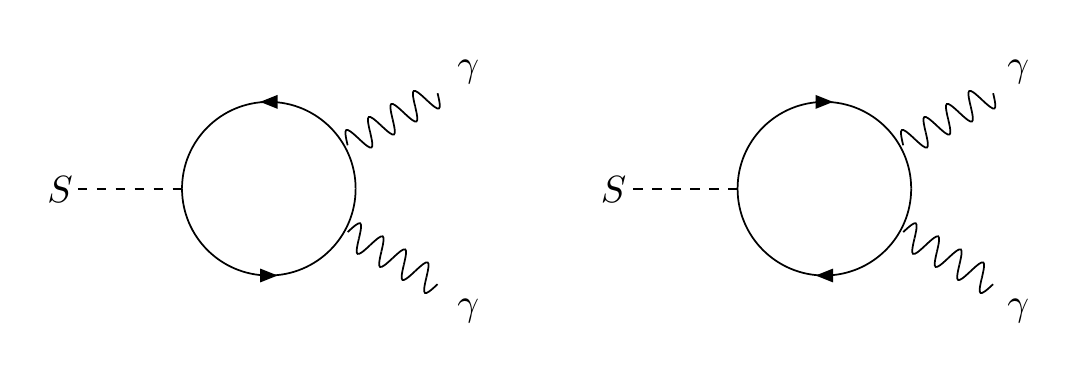}
\caption{\label{sgg}\footnotesize
$S\mbox{-}\gamma\mbox{-}\gamma$ coupling}
\end{figure}

\vspace{0.5cm}
\noindent
$\bullet$\underline{$S\mbox{-}\gamma\mbox{-}\gamma$ amplitude}\\
The next example is the $S(p)\mbox{-}\gamma(k)\mbox{-}\gamma(k')$ 
three-point function.
There are two diagrams at the one-loop level as shown in Fig.~\ref{sgg}.
Again using the usual technique, we obtain
the unsubtracted amplitude
\be
{\cal A}_{\mu\nu}(\Lambda;k,k')
&=&i\frac{e^2 Q^2 y N_c N_f M}{4\pi^2}\int_0^1 dx\int_0^{1-x} dy~ \frac{2\Lambda^4}{(\Lambda^2+D^2)^2}\times\nn\\
& &
\left[~g_{\mu\nu} -
\frac{(1-4 x y)(~g_{\mu\nu}~k\cdot k'-k_\mu k'_\nu~)}{D^2}
~\right]~,
\label{sgg1}
\ee
where
\be
D^2 =M^2-2 x y~  k\cdot k'~.
\label{D}
\ee
The last term in the square bracket
has an imaginary part and gauge invariant.
The first $g_{\mu\nu}$ term in the square bracket
is not gauge invariant.
It is obvious  that least subtraction procedure
implies a complete subtraction of this term.
The subtracted amplitude is
\begin{align}
{\cal A}^R_{\mu\nu}(\Lambda;k,k')
&=&-i\frac{e^2 Q^2 y N_c N_f  M}{4\pi^2}\int_0^1 dx\int_0^{1-x} dy~\frac{2\Lambda^4}{(\Lambda^2+D^2)^2}
\frac{(1-4 x y)(~g_{\mu\nu}~k\cdot k'-k_\mu k'_\nu~)}{D^2}. \displaybreak[0]
\end{align}

\vspace{0.5cm}
\noindent
$\bullet$\underline{$\phi\mbox{-}\phi\mbox{-}\gamma\mbox{-}\gamma$ amplitude}\\
The next example is the 
$\phi(p)\mbox{-}\phi(p')\mbox{-}\gamma(k)\mbox{-}\gamma(k')$ four-point 
function.
The diagrams at the one-loop level are shown 
in Fig.~\ref{dmdmgg5}$\sim$ \ref{dmdmgg3}
with the sum of the unsubtracted  amplitudes:
\be
{\cal A}_{\mu\nu}(\Lambda;k,k',p,p')
&=&{\cal A}_{\mu\nu}^{(a)}(\Lambda;k,k',p,p')
+{\cal A}_{\mu\nu}^{(b)}(\Lambda;k,k',p,p')
+{\cal A}_{\mu\nu}^{(c)}(\Lambda;k,k',p,p')~.
\label{Amunu1}
\ee
The diagrams of Fig.~\ref{dmdmgg5}, which give  
${\cal A}_{\mu\nu}^{(a)}$ in (\ref{Amunu1}), are simpler to compute,
because the structure is the same as Fig.~\ref{sgg}. 
Again using the usual technique, we obtain
the unsubtracted amplitude (with $N_f=3$)
\be
{\cal A}_{\mu\nu}^{(a)}(\Lambda;k,k')
&=&-i\left[\frac{e^2  Q^2 N_c }{2\pi^2}\right]
\left[ \frac{G_D M}{8G^2}\right]
\int_0^1 dx\int_0^{1-x} dy\nn\\
& & \times \frac{2\Lambda^4}{  (\Lambda^2+D^2)^2}\left[
g_{\mu\nu}
-\frac{(1-4 x y)(~g_{\mu\nu}~k\cdot k'-k_\mu k'_\nu~)}
{D^2}\right]~,
\ee
where $D^2$ is given in (\ref{D}).
The non-gauge invariant terms have the same structure as
(\ref{sgg1}). Therefore, we subtract the non-gauge invariant
$g_{\mu\nu}$ term:
\be
{\cal A}_{\mu\nu}^{R(a)} &=&
(~g_{\mu\nu}~k\cdot k'-k_\mu k'_\nu~){\cal A}^{R(a)}
=
{\cal A}_{\mu\nu}^{(a)}~\mbox{with the first}~
g_{\mu\nu} ~\mbox{term omitted.}
\label{ARa}
\ee

The tensor structure of the other diagrams is more complicated.
The amplitude has to satisfy the Bose symmetry:
\be
{\cal A}_{\mu\nu}(\Lambda;k,k',p,p') &=&
{\cal A}_{\nu\mu}(\Lambda;k',k,p,p') ~
\mbox{and}~{\cal A}_{\mu\nu}(\Lambda;k,k',p,p') =
{\cal A}_{\mu\nu}(\Lambda;k,k',p',p) ~.
\label{bose}
\ee
To proceed
 we introduce the Mandelstam variables:
\be
S&=& (p+p')^2=2 m_{\rm DM}^2+2 p\cdot p'
= (k+k')^2=2k\cdot k'~,\\~
T &=& (p-k)^2= m_{\rm DM}^2-2 p\cdot k~,~
U = (p-k')^2= m_{\rm DM}^2-2 p\cdot k'~.
\ee
All the dot products of the momenta and $m_{\rm DM}^2$
can be expressed as a function of $S,T$ and $U$.
The most general tensor structure, which is consistent with
the Bose symmetry (\ref{bose}) is\footnote{Because of the on-shell gauge invariance,
we have suppressed terms proportional to $k_\nu$ and
$k'_\mu$ in (\ref{Tensor}).}
\begin{align}
{\cal A}^{(b+c)}_{\mu\nu}(\Lambda;k,k',p,p') =& {\cal A}^{(b)}_{\mu\nu}(\Lambda;k,k',p,p')+{\cal A}^{(c)}_{\mu\nu}(\Lambda;k,k',p,p') \displaybreak[0] \nn\\
=& i\left[\frac{e^2 Q^2 N_c}{2\pi^2}\right]\left( 1-\frac{G_D \langle \sigma\rangle}{8G^2}\right)^2\left\{g_{\mu\nu}~{\cal A}_g(\Lambda;S,T,U)\right. \displaybreak[0]\nn\\
&+k_\mu k'_\nu ~{\cal A}_{kk}(\Lambda;S,T,U)+(p_\mu p_\nu +p'_\mu p'_\nu )~{\cal A}_{p}(\Lambda;S,T,U)\nn\\
&+p'_\mu p_\nu ~{\cal A}_{p'}(\Lambda;S,T,U)+p_\mu p'_\nu~ {\cal A}_{p'}(\Lambda;S,U,T)\nn \\
&\left. +(k_\mu p'_\nu+p_\mu k'_\nu ){\cal A}_{k}(\Lambda;S,T,U)+(k_\mu p_\nu+p'_\mu k'_\nu ){\cal A}_{k}(\Lambda;S,U,T)\right\},
\label{Tensor}
\end{align}
where the amplitudes ${\cal A}_{i}$'s have to satisfy 
\be
{\cal A}_{g,kk,p}(\Lambda;S,T,U) &=&
{\cal A}_{g,kk,p}(\Lambda;S,U,T)~.
\label{bose1}
\ee
Gauge invariance means that the following quantities vanish:
\be
k^\nu {\cal A}_{\mu\nu}(\Lambda;k,k',p,p') &=&
k_\mu ~{\cal G}_1(\Lambda;S,T,U)+p_\mu
~{\cal G}_2(\Lambda;S,T,U)  +p'_\mu ~{\cal G}_3(\Lambda;S,T,U),
\label{kA1}\\
k'^\mu{\cal A}_{\mu\nu}(\Lambda;k,k',p,p') &=&
k'_\nu~ {\cal G}'_1(\Lambda;S,T,U)+p_\nu
~{\cal G}'_2(\Lambda;S,T,U)  +p'_\nu~ {\cal G}'_3(\Lambda;S,T,U),
\label{kA2}
\ee
where we impose the on-shell condition
$k^2=k'^2=0, p^2=p'^2=m_{\rm DM}^2$ along
with the four momentum conservation
$p+p'=k+k'$, and ${\cal G}_i$'s are 
defined as
\begin{align}
&{\cal G}_1(\Lambda;S,T,U) ={\cal G}'_1(\Lambda;S,T,U) \nn\\
=&{\cal A}_g(\Lambda;S,T,U)+\frac{S}{4}\left[ 2{\cal A}_{kk}(\Lambda;S,T,U) +{\cal A}_{k}(\Lambda;S,T,U)+{\cal A}_{k}(\Lambda;S,U,T)\right]\nn\\
&+\frac{1}{4}(T-U)\left[{\cal A}_{k}(\Lambda;S,T,U)-{\cal A}_{k}(\Lambda;S,U,T)\right]~,\label{G1}
\end{align}
\begin{align}
&{\cal G}_2(\Lambda;S,T,U)={\cal G}'_3(\Lambda;S,T,U)={\cal G}_3(\Lambda;S,U,T) \nn\\
=&\frac{1}{4}(S-T+U){\cal A}_{p}(\Lambda;S,T,U)+\frac{S}{2}{\cal A}_{k}(\Lambda;S,T,U)+\frac{1}{4}(S+T-U){\cal A}_{p'}(\Lambda;S,U,T),
\end{align}
\begin{align}
&{\cal G}_3(\Lambda;S,T,U)={\cal G}'_2(\Lambda;S,T,U) ={\cal G}_2(\Lambda;S,U,T)\label{G2}\nn \\
=&\frac{1}{4}(S+T-U){\cal A}_{p}(\Lambda;S,U,T)+ \frac{S}{2}{\cal A}_{k}(\Lambda;S,U,T)+\frac{1}{4}(S-T+U){\cal A}_{p'}(\Lambda;S,T,U).
\end{align}
Gauge invariance requires that all ${\cal G}_i$'s should vanish identically.
We have calculated them 
for ${\cal A}_{\mu\nu}^{(b+c)}={\cal A}_{\mu\nu}^{(b)}
+{\cal A}_{\mu\nu}^{(c)}$ explicitly for a small external momenta
$\sim \mathcal{O}(\epsilon)$ and 
find
\begin{align}
&\epsilon=0 : {\cal G}_1 = \frac{\Lambda^4}{(\Lambda^2+M^2)^2},\quad {\cal G}_2 =0~,\label{G10} \\ \displaybreak[0]
&\epsilon^2 :{\cal G}_1 =\frac{11\Lambda^4}{40(\Lambda^2+M^2)^3}\left[S+2(T+U)\right],\;{\cal G}_2 =\frac{3\Lambda^4}{40(\Lambda^2+M^2)^3}(T-U)~,
\end{align}
\begin{align}
\epsilon^4 :{\cal G}_1 =&\frac{\Lambda^4}{280(\Lambda^2+M^2)^4}\left[82S^2+275 S(T+U)+64\left\{4(T^2+U^2)+7 TU\right\}\right]~,\nn\\
{\cal G}_2 =&\frac{3\Lambda^4}{280(\Lambda^2+M^2)^4}(T-U)\left[16S+13(T+U)\right]~,
\end{align}
\begin{align}
\epsilon^6 :{\cal G}_1 =&\frac{\Lambda^4}{1008(\Lambda^2+M^2)^5}\left[517S^3+2168 S^2(T+U)+10 S\left\{337(T^2+U^2)+586 TU\right\}\nn\right.\\
&\left.+3\left\{761 (T^3+U^3)+1591(T^2U+T U^2)\right\}\right]~,\\
{\cal G}_2 =&\frac{\Lambda^4}{336(\Lambda^2+M^2)^5}(T-U)\left[116S^2+239S(T+U)+17\left\{7(T^2+U^2)+10TU\right\}\right]~,\nn
\end{align}
\begin{align}
\epsilon^8 :{\cal G}_1 =&\frac{\Lambda^4}{1584(\Lambda^2+M^2)^6}\left[2142S^4+10058 S^3(T+U)+2S^2(\left\{9701(T^2+U^2)+17284 TU\right\}\nn\right.\\
&+15 S\left\{1399 (T^3+U^3)+3137(T^2U+T U^2)\right\}+3 \left\{3993(T^4+U^4)\right.\nn\\
&\left.\left.+9178(T^3 U+T U^3)+11098 T^2 U^2\right\}\right]~,\nn\\
{\cal G}_2 =&\frac{\Lambda^4}{1584(\Lambda^2+M^2)^6}(T-U)\left[1512S^3+4007S^2 (T+U)\right.\nn\\
& \left.+15S\left\{319(T^2+U^2)+512TU\right\} +630\left\{3(T^3+U^3)+5(T^2U+TU^2)\right\}\right]~.
\label{G28}
\end{align}
As we see from (\ref{G10}) $\sim$ (\ref{G28}) that the non-gauge invariant function $ {\cal G}_1$
has the same structure as $ {\cal G}$ in (\ref{calG})
for the  photon self-energy. Therefore, we subtract this term
from the amplitude so that the function ${\cal A}_{g}$ is replaced
by
\be
{\cal A}^R_{g} &=& {\cal A}_{g}-{\cal G}_1~,
\label{ARg}
\ee
where  ${\cal A}_{g}$ and ${\cal G}_1$ are defined in
 (\ref{Tensor}) and (\ref{G1}),
respectively.
At $\mathcal{O}(\epsilon^2)$ 
${\cal G}_2$ becomes non-zero. As we see from
(\ref{kA1}) and (\ref{kA2}), this non-gauge invariant term
requires an introduction of a new tensor structure for counter terms.
We see from (\ref{G10}) $\sim$ (\ref{G28}) that 
$ {\cal G}_2$ is proportional to
$(T-U)$, so that we can rewrite it
as 
\be
{\cal G}_2(\Lambda;S,T,U)&=&(T-U) \hat{{\cal G}}_2
(\Lambda;S,T,U)~,
\label{G2hat}
\ee
which  will be justified below to all orders in the
expansion of $k'$s and $p'$s. 
Therefore, we can cancel this non-gauge
invariant term by adding counter terms such that
${\cal A}_{p}$ and ${\cal A}_{p'}$ change according to
\be
{\cal A}_{p} &\to & {\cal A}^R_{p}={\cal A}_{p}+2\hat{{\cal G}}_2~,~
{\cal A}_{p'} \to {\cal A}^R_{p'}={\cal A}_{p'}-2\hat{{\cal G}}_2~,
\label{ARp}
\ee
where ${\cal A}_{p}$ and  ${\cal A}_{p'}$ are defined in (\ref{Tensor}). 
Since ${\cal A}_{p} $ has to satisfy the Bose symmetry (\ref{bose1}),
$\hat{{\cal G}}_2$, too,  has to satisfy the same symmetry.
We have numerically checked that
\be
{\cal G}_2(\Lambda;S,T,T)&=&0~,~
{\cal G}_2(\Lambda;S,T,U)=-{\cal G}_2(\Lambda;S,U,T)~
\label{G2-1}
\ee
is satisfied within an accuracy that we can get, which implies that $ 
\hat{{\cal G}}_2(\Lambda;S,T,U)=
\hat{{\cal G}}_2(\Lambda;S,U,T) $.

Finally, we have the gauge invariant 
$\phi\mbox{-}\phi\mbox{-}\gamma\mbox{-}\gamma$ four-point function:
\begin{align}
{\cal A}^R_{\mu\nu}(\Lambda;k,k',p,p')=&{\cal A}^{R(a)}_{\mu\nu}(\Lambda;k,k',p,p')+i\left[\frac{ e^2 Q^2 N_c}{2\pi^2}\right]
\left( 1-\frac{G_D\langle \sigma\rangle}{8G^2}\right)^2 \left\{g_{\mu\nu}~{\cal A}^R_g(\Lambda;S,T,U)\right.\nn\\
&+k_\mu k'_\nu ~{\cal A}_{kk}(\Lambda;S,T,U)+(p_\mu p_\nu +p'_\mu p'_\nu )~{\cal A}^R_{p}(\Lambda;S,T,U)\nn\\
&+
p'_\mu p_\nu~ {\cal A}^R_{p'}(\Lambda;S,T,U)
+p_\mu p'_\nu ~{\cal A}^R_{p'}(\Lambda;S,U,T)\nn\\
&\left. +(k_\mu p'_\nu+p_\mu k'_\nu )~{\cal A}_{k}(\Lambda;S,T,U)+
 (k_\mu p_\nu+p'_\mu k'_\nu )~
 {\cal A}_{k}(\Lambda;S,U,T)\right\}~,
 \label{AmunuR}~
\end{align}
where ${\cal A}^{R(a)}$ is given in (\ref{ARa}).
Note that for the case $T=U$ the function ${\cal G}_2$ vanishes.
In this case, therefore, all the counter terms are
proportional to the metric tensor.
In the center of mass system, $T=U$ implies $p=p'=
(m_{\rm DM}, {\bf 0})$, and only $k$ and $k'$
are independent Lorentz vectors ($p_\mu=(k_\mu+k'_\mu)/2$).
In this system, the non-gauge invariant terms of ${\cal A}_{\mu\nu}$
are proportional to $g_{\mu\nu}$.
No Lorentz transformation can produce non-gauge
invariant terms proportional to $k_\mu p_\nu$ or
$p_\mu p_\nu$ from the $g_{\mu\nu}$ term.
This is the reason why ${\cal G}_2$ vanishes when  $T=U$, and
therefore the assumption that this function takes the form
${\cal G}_2=(T-U) \hat{{\cal G}}_2$ is justified
to all orders in the expansion of $k'$s and $p'$s.
The amplitude with $p=p'=
(m_{\rm DM}, {\bf 0})$ will be used for the annihilation
of two DM's into two photons in sect. V.

The tensor structure (\ref{Tensor}) is the most general one.
However, using the on-shell conditions
$k+k'=p+p'$ we can eliminate
$k$ and $k'$, because $k_\nu$ and $k'_\mu$ terms
do not  contribute neither to the physical amplitude
nor to the on-shell gauge invariance conditions (\ref{kA1}) and (\ref{kA2})
and hence can be suppressed as we have done in (\ref{Tensor}). 
In this basis, ${\cal A}^{(b+c)}_{\mu\nu}(\Lambda;k,k',p,p')$
of (\ref{Tensor}) becomes
\be
{\cal A}^{(b+c)}_{\mu\nu}(\Lambda;k,k',p,p') 
& =&
i\left[\frac{e^2 Q^2 N_c}{2\pi^2}\right]
\left( 1-\frac{G_D\langle \sigma\rangle}{8G^2}\right)^2\nn\\
& &\times\left\{
g_{\mu\nu}~{\cal A}_g(\Lambda;S,T,U)+
(p_\mu p_\nu +p'_\mu p'_\nu )~\tilde{{\cal A}}_{p}(\Lambda;S,T,U)
\right.\nn\\
& &\left.+
p'_\mu p_\nu ~\tilde{{\cal A}}_{p'}(\Lambda;S,T,U)
+p_\mu p'_\nu~ \tilde{{\cal A}}_{p'}(\Lambda;S,U,T)\right\}
\label{tensor1}~,
\ee
where
\be
\tilde{{\cal A}}_{p}(\Lambda;S,T,U)
&=&
{\cal A}_{p}(\Lambda;S,T,U)+
{\cal A}_{kk}(\Lambda;S,T,U)+
{\cal A}_{k}(\Lambda;S,T,U)\nn\\
& &+{\cal A}_{k}(\Lambda;S,U,T)~,\\
\tilde{{\cal A}}_{p'}(\Lambda;S,T,U)
&=&
{\cal A}_{p'}(\Lambda;S,T,U)+
{\cal A}_{kk}(\Lambda;S,T,U)+
2{\cal A}_{k}(\Lambda;S,U,T)~.
\ee
Then the gauge invariance conditions (\ref{kA1}) and (\ref{kA2})
mean that
\be
\tilde{{\cal G}}_2(\Lambda;S,T,U) &=&\tilde{{\cal G}}_3(\Lambda;S,U,T)
 \nn\\
&=&\frac{1}{4}(S-T+U)\tilde{{\cal A}}_{p}(\Lambda;S,T,U)+
\frac{1}{4}(S+T-U)\tilde{{\cal A}}_{p'}(\Lambda;S,U,T)
\label{G2tilde1}\\
&= &{\cal G}_1(\Lambda;S,T,U) 
+(T-U)\hat{{\cal G}}_2(\Lambda;S,T,U)~
\label{G2tilde2}
\ee
has to vanish, where ${\cal G}_1$ and $\hat{{\cal G}}_2$
are defined in (\ref{G1}) and (\ref{G2hat}), respectively.
From (\ref{G2tilde1}) and (\ref{G2tilde2}) it is now  obvious
how to restore gauge invariance:
From (\ref{G2tilde2}) we see that $\tilde{{\cal G}}_2(\Lambda;S,T,U)$
can be uniquely divided into  the even
 and odd part under the interchange $T\leftrightarrow U$,
 because ${\cal G}_1$ and $\hat{{\cal G}}_2$ are even functions.
 The odd part can be canceled by
 \be
\tilde{{\cal A}}_{p} &\to & \tilde{{\cal A}}^R_{p}=
\tilde{{\cal A}}_{p}+2\hat{{\cal G}}_2~,~
\tilde{{\cal A}}_{p'} \to \tilde{{\cal A}}^R_{p'}=
\tilde{{\cal A}}_{p'}-2\hat{{\cal G}}_2~,
\label{ARp2}
\ee
as  in the case of ${\cal A}_{p}$ and
${\cal A}_{p'}$   (see (\ref{ARp})). 
The even part can be canceled by the redefinition  of
$A_g$ which is defined  in  (\ref{ARg}).
The resulting gauge invariant subtracted amplitude 
is identical with  (\ref{AmunuR}) up to on-shell conditions.


\begin{thebibliography}{99}
 \bibitem{Aad:2012tfa}
  G.~Aad {\it et al.}  [ATLAS Collaboration],
  Phys.\ Lett.\ B {\bf 716} (2012) 1
  [arXiv:1207.7214 [hep-ex]].
  
  \bibitem{Chatrchyan:2012ufa}
  S.~Chatrchyan {\it et al.}  [CMS Collaboration],
  Phys.\ Lett.\ B {\bf 716} (2012) 30
  [arXiv:1207.7235 [hep-ex]].
  

  
  \bibitem{Holthausen:2011aa}
  M.~Holthausen, K.~S.~Lim and M.~Lindner,
  JHEP {\bf 1202} (2012) 037
  [arXiv:1112.2415 [hep-ph]].
  \bibitem{Degrassi:2012ry}
  G.~Degrassi, S.~Di Vita, J.~Elias-Miro, J.~R.~Espinosa, G.~F.~Giudice, G.~Isidori and A.~Strumia,
  JHEP {\bf 1208} (2012) 098
  [arXiv:1205.6497 [hep-ph]];
  D.~Buttazzo, G.~Degrassi, P.~P.~Giardino, G.~F.~Giudice, F.~Sala, A.~Salvio and A.~Strumia,
  JHEP {\bf 1312} (2013) 089
  [arXiv:1307.3536].
  \bibitem{Bezrukov:2012sa}
  F.~Bezrukov, M.~Y.~Kalmykov, B.~A.~Kniehl and M.~Shaposhnikov,
  JHEP {\bf 1210} (2012) 140
  [arXiv:1205.2893 [hep-ph]].
  
    \bibitem{Kubo:2014ova}
  J.~Kubo, K.~S.~Lim and M.~Lindner,
  arXiv:1403.4262 [hep-ph].
 
  
 \bibitem{Coleman:1973jx}
    S.~R.~Coleman and E.~J.~Weinberg,
  Phys.\ Rev.\ D {\bf 7} (1973) 1888.

  \bibitem{Fatelo:1994qf}
  J.~P.~Fatelo, J.~M.~Gerard, T.~Hambye and J.~Weyers,
  Phys.\ Rev.\ Lett.\  {\bf 74} (1995) 492.
  
  \bibitem{Hempfling:1996ht}
  R.~Hempfling,
  Phys.\ Lett.\ B {\bf 379} (1996) 153
  [hep-ph/9604278].
  
  \bibitem{Hambye:1995fr}
  T.~Hambye,
  Phys.\ Lett.\ B {\bf 371} (1996) 87
  [hep-ph/9510266].
 
  \bibitem{Meissner:2006zh}
  K.~A.~Meissner and H.~Nicolai,
  Phys.\ Lett.\ B {\bf 648} (2007) 312
  [hep-th/0612165];
  K.~A.~Meissner and H.~Nicolai,
  Phys.\ Lett.\ B {\bf 660} (2008) 260
  [arXiv:0710.2840 [hep-th]];
  K.~A.~Meissner and H.~Nicolai,
  Phys.\ Rev.\ D {\bf 80} (2009) 086005
  [arXiv:0907.3298 [hep-th]].
 
  \bibitem{Foot:2007as}
  R.~Foot, A.~Kobakhidze and R.~R.~Volkas,
  Phys.\ Lett.\ B {\bf 655} (2007) 156
  [arXiv:0704.1165 [hep-ph]];
  Phys.\ Rev.\ D {\bf 84} (2011) 075010
  [arXiv:1012.4848 [hep-ph]].
  
  
   \bibitem{Foot:2007ay}
  R.~Foot, A.~Kobakhidze, K.~.L.~McDonald and R.~.R.~Volkas,
  Phys.\ Rev.\ D {\bf 76} (2007) 075014
  [arXiv:0706.1829 [hep-ph]];
  Phys.\ Rev.\ D {\bf 77} (2008) 035006
  [arXiv:0709.2750 [hep-ph]];
  arXiv:1310.0223 [hep-ph].
      
\bibitem{Chang:2007ki}
  W.~-F.~Chang, J.~N.~Ng and J.~M.~S.~Wu,
  Phys.\ Rev.\ D {\bf 75} (2007) 115016
  [hep-ph/0701254 [HEP-PH]].
  
    \bibitem{Hambye:2007vf}
  T.~Hambye and M.~H.~G.~Tytgat,
  Phys.\ Lett.\ B {\bf 659} (2008) 651
  [arXiv:0707.0633 [hep-ph]].
  
  \bibitem{Iso:2009ss}
  S.~Iso, N.~Okada and Y.~Orikasa,
  Phys.\ Lett.\ B {\bf 676} (2009) 81
  [arXiv:0902.4050 [hep-ph]];
  Phys.\ Rev.\ D {\bf 80} (2009) 115007
  [arXiv:0909.0128 [hep-ph]];
  PTEP {\bf 2013} (2013) 023B08
  [arXiv:1210.2848 [hep-ph]].

  \bibitem{Holthausen:2009uc}
  M.~Holthausen, M.~Lindner and M.~A.~Schmidt,
  Phys.\ Rev.\ D {\bf 82} (2010) 055002
  [arXiv:0911.0710 [hep-ph]].
  
\bibitem{Ishiwata:2011aa} 
  K.~Ishiwata,
  Phys.\ Lett.\ B {\bf 710}, 134 (2012)
  [arXiv:1112.2696 [hep-ph]].
  
  \bibitem{Khoze:2013uia}
  V.~V.~Khoze,
  JHEP {\bf 1311} (2013) 215
  [arXiv:1308.6338 [hep-ph]].
    
    \bibitem{Kawamura:2013kua}
  Y.~Kawamura,
  PTEP {\bf 2013} (2013) 11,  113B04
  [arXiv:1308.5069 [hep-ph]].
  
  \bibitem{Gretsch:2013ooa}
  F.~Gretsch and A.~Monin,
  arXiv:1308.3863 [hep-th].
  
  \bibitem{Carone:2013wla}
  C.~D.~Carone and R.~Ramos,
  Phys.\ Rev.\ D {\bf 88} (2013) 055020
  [arXiv:1307.8428 [hep-ph]].
  
  \bibitem{Khoze:2013oga}
  V.~V.~Khoze and G.~Ro,
  JHEP {\bf 1310} (2013) 075
  [arXiv:1307.3764].

\bibitem{Gabrielli:2013hma} 
  E.~Gabrielli, M.~Heikinheimo, K.~Kannike, A.~Racioppi, M.~Raidal and C.~Spethmann,
  Phys.\ Rev.\ D {\bf 89}, 015017 (2014)
  [arXiv:1309.6632 [hep-ph]].
  
  \bibitem{Englert:2013gz}
  C.~Englert, J.~Jaeckel, V.~V.~Khoze and M.~Spannowsky,
  JHEP {\bf 1304} (2013) 060
  [arXiv:1301.4224 [hep-ph]].
  
  \bibitem{Farzinnia:2013pga}
  A.~Farzinnia, H.~-J.~He and J.~Ren,
  Phys.\ Lett.\ B {\bf 727} (2013) 141
  [arXiv:1308.0295 [hep-ph]].
  
    
  \bibitem{Abel:2013mya}
  S.~Abel and A.~Mariotti,
  arXiv:1312.5335 [hep-ph].
  
    \bibitem{Ibe:2013rpa}
  M.~Ibe, S.~Matsumoto and T.~T.~Yanagida,
  arXiv:1312.7108 [hep-ph].
  
  \bibitem{Hill:2014mqa}
  C.~T.~Hill,
  arXiv:1401.4185 [hep-ph].
  
   \bibitem{Guo:2014bha}
  J.~Guo and Z.~Kang,
  arXiv:1401.5609 [hep-ph].
  
    \bibitem{Radovcic:2014rea}
  B.~Radovcic and S.~Benic,
  arXiv:1401.8183 [hep-ph].
  
  \bibitem{Khoze:2014xha}
  V.~V.~Khoze, C.~McCabe and G.~Ro,
  arXiv:1403.4953 [hep-ph].
  
\bibitem{Davoudiasl:2014pya}
    H.~Davoudiasl and I.~M.~Lewis,
   arXiv:1404.6260 [hep-ph].
  
  \bibitem{Chankowski:2014fva}
  P.~H.~Chankowski, A.~Lewandowski, K.~A.~Meissner and H.~Nicolai,
  arXiv:1404.0548 [hep-ph].

\bibitem{Callan:1970yg}
  C.~G.~Callan, Jr.,
  Phys.\ Rev.\ D {\bf 2} (1970) 1541;
  K.~Symanzik,
  Commun.\ Math.\ Phys.\  {\bf 18} (1970) 227.

\bibitem{Bardeen:1995kv} 
  W.~A.~Bardeen,
  FERMILAB-CONF-95-391-T.

\bibitem{Hur:2007uz}
  T.~Hur, D.~-W.~Jung, P.~Ko and J.~Y.~Lee,
  Phys.\ Lett.\ B {\bf 696} (2011) 262
  [arXiv:0709.1218 [hep-ph]].
\bibitem{Hur:2011sv}
  T.~Hur and P.~Ko,
  Phys.\ Rev.\ Lett.\  {\bf 106} (2011) 141802
  [arXiv:1103.2571 [hep-ph]].

\bibitem{Heikinheimo:2013fta}
  M.~Heikinheimo, A.~Racioppi, M.~Raidal, C.~Spethmann and K.~Tuominen,
  arXiv:1304.7006 [hep-ph].
  
\bibitem{Holthausen:2013ota}
  M.~Holthausen, J.~Kubo, K.~S.~Lim and M.~Lindner,
  arXiv:1310.4423 [hep-ph].
  
    
  \bibitem{Strassler:2006im}
  M.~J.~Strassler and K.~M.~Zurek,
  Phys.\ Lett.\ B {\bf 651} (2007) 374
  [hep-ph/0604261];
  T.~Han, Z.~Si, K.~M.~Zurek and M.~J.~Strassler,
  JHEP {\bf 0807} (2008) 008
  [arXiv:0712.2041 [hep-ph]].
    \bibitem{Bringmann:2007nk}
  T.~Bringmann, L.~Bergstrom and J.~Edsjo,
  JHEP {\bf 0801} (2008) 049
  [arXiv:0710.3169 [hep-ph]].
  

\bibitem{Bertone:2009cb}
  G.~Bertone, C.~B.~Jackson, G.~Shaughnessy, T.~M.~P.~Tait and A.~Vallinotto,
  Phys.\ Rev.\ D {\bf 80} (2009) 023512
  [arXiv:0904.1442 [astro-ph.HE]].
  
\bibitem{Laha:2012fg} 
  R.~Laha, K.~C.~Y.~Ng, B.~Dasgupta and S.~Horiuchi,
  Phys.\ Rev.\ D {\bf 87}, no. 4, 043516 (2013)
  [arXiv:1208.5488 [astro-ph.CO]].


\bibitem{Ackermann:2012qk}
  M.~Ackermann {\it et al.}  [LAT Collaboration],
  Phys.\ Rev.\ D {\bf 86} (2012) 022002
  [arXiv:1205.2739 [astro-ph.HE]].
  
\bibitem{Gustafsson:2013fca}
  M.~Gustafsson [ for the Fermi-LAT Collaboration],
  arXiv:1310.2953 [astro-ph.HE].


  \bibitem{Abramowski:2013ax}
  A.~Abramowski {\it et al.}  [H.E.S.S. Collaboration],
  Phys.\ Rev.\ Lett.\  {\bf 110} (2013) 041301
  [arXiv:1301.1173 [astro-ph.HE]].
  \bibitem{Bulbul:2014sua}
  E.~Bulbul, M.~Markevitch, A.~Foster, R.~K.~Smith, M.~Loewenstein and S.~W.~Randall,
  arXiv:1402.2301 [astro-ph.CO].
  
  \bibitem{Boyarsky:2014jta} 
  A.~Boyarsky, O.~Ruchayskiy, D.~Iakubovskyi and J.~Franse,
  arXiv:1402.4119 [astro-ph.CO].
\bibitem{Dolgov:2013una} 
  A.~D.~Dolgov, S.~L.~Dubovsky, G.~I.~Rubtsov and I.~I.~Tkachev,
  Phys.\ Rev.\ D {\bf 88}, no. 11, 117701 (2013)
  [arXiv:1310.2376 [hep-ph]].
\bibitem{Langacker:2011db} 
  P.~Langacker and G.~Steigman,
  Phys.\ Rev.\ D {\bf 84}, 065040 (2011)
  [arXiv:1107.3131 [hep-ph]].
  
   \bibitem{Nambu:1960xd}
  Y.~Nambu,
  Phys.\ Rev.\ Lett.\  {\bf 4} (1960) 380.
 
\bibitem{Nambu:1961tp}
  Y.~Nambu and G.~Jona-Lasinio,
  Phys.\ Rev.\  {\bf 122} (1961) 345;
  Phys.\ Rev.\  {\bf 124} (1961) 246.
  
  
  \bibitem{Klevansky:1992qe}
  S.~P.~Klevansky,
  Rev.\ Mod.\ Phys.\  {\bf 64} (1992) 649.

\bibitem{Hatsuda:1994pi}
  T.~Hatsuda and T.~Kunihiro,
  Phys.\ Rept.\  {\bf 247} (1994) 221
  [hep-ph/9401310].

 
 \bibitem{Kunihiro:1983ej}
  T.~Kunihiro and T.~Hatsuda,
  Prog.\ Theor.\ Phys.\  {\bf 71} (1984) 1332;
  Phys.\ Rev.\ Lett.\  {\bf 55}, 158 (1985);
  Phys.\ Lett.\ B {\bf 206} (1988) 385
   [Erratum-ibid.\  {\bf 210} (1988) 278].
  
  \bibitem{Lowenstein:1975rf}
  J.~H.~Lowenstein and W.~Zimmermann,
  Commun.\ Math.\ Phys.\  {\bf 46} (1976) 105;
  Commun.\ Math.\ Phys.\  {\bf 44} (1975) 73
   [Lect.\ Notes Phys.\  {\bf 558} (2000) 310].
  \bibitem{Poggio:1976qr}
  E.~C.~Poggio and H.~R.~Quinn,
  Phys.\ Rev.\ D {\bf 14} (1976) 578.

  
    
  \bibitem{Ade:2013ktc}
  P.~A.~R.~Ade {\it et al.}  [Planck Collaboration],
  arXiv:1303.5062 [astro-ph.CO].
  
    \bibitem{Aprile:2012nq}
  E.~Aprile {\it et al.}  [XENON100 Collaboration],
  Phys.\ Rev.\ Lett.\  {\bf 109} (2012) 181301
  [arXiv:1207.5988 [astro-ph.CO]].
  
  \bibitem{Akerib:2013tjd}
  D.~S.~Akerib {\it et al.}  [LUX Collaboration],
  arXiv:1310.8214 [astro-ph.CO].
  
  \bibitem{Aprile:2012zx}
  E.~Aprile [XENON1T Collaboration],
  arXiv:1206.6288 [astro-ph.IM].
  



  
  \bibitem{Jauch}
  J. M.~Jauch and F.~Rohrlich,
 ``The Theory of Photons and Electrons'',
Addison-Wesley Pub. Company, Inc. 1959 


\bibitem{Suganuma:1990nn}
  H.~Suganuma and T.~Tatsumi,
  Annals Phys.\  {\bf 208} (1991) 470.

  \bibitem{Klevansky:1991ey}
  S.~P.~Klevansky, J.~Janicke and R.~H.~Lemmer,
  Phys.\ Rev.\ D {\bf 43} (1991) 3040.

  
  \bibitem{Tulin:2012uq}
  S.~Tulin, H.~-B.~Yu and K.~M.~Zurek,
  Phys.\ Rev.\ D {\bf 87} (2013) 036011
  [arXiv:1208.0009 [hep-ph]].

  
  \bibitem{Kanemura:2011nm}
  S.~Kanemura, S.~Matsumoto, T.~Nabeshima and H.~Taniguchi,
  Phys.\ Lett.\ B {\bf 701} (2011) 591
  [arXiv:1102.5147 [hep-ph]].
    \end{thebibliography}
\end{document}